\documentclass{aa}  

\usepackage{graphicx}
\usepackage{txfonts}

\usepackage{graphicx,rotating,amssymb,amsmath}
\usepackage{booktabs}
\usepackage{subfig}
\usepackage{float,capt-of}
\usepackage{natbib}
\usepackage{multirow}
\usepackage[english]{babel}
\usepackage{morefloats}
\usepackage{color}
\usepackage{xcolor}

\usepackage{enumitem}
\usepackage{hyperref}
\usepackage{multirow}
\usepackage{multicol}
\usepackage{ulem}

\newcommand{\acounits}{\mathrm{(K\,km\,s^{-1}\,pc^2)^{-1}}}
\newcommand{\fmol}{f_\mathrm{mol}}
\newcommand{\tmol}{\tau_\mathrm{mol}}
\newcommand{\fmhi}{f_\mathrm{HI}}
\newcommand{\mmol}{M_\mathrm{mol}}
\newcommand{\sfr}{\mathrm{SFR}}
\newcommand{\msun}{M_\odot}
\newcommand{\bx}{} 
\newcommand{\bxx}{} 

\begin{document}
   \title{Molecular gas scaling relations for local star-forming galaxies in the low-$M_*$ regime}

   \author{B. Hagedorn \inst{1}
          \and
          C. Cicone \inst{1} 
          \and
          M. Sarzi \inst{2}
          \and
          A. Saintonge \inst{3}
          \and
          P. Severgnini \inst{4}
          \and
          C. Vignali \inst{5,6}
          \and
          S. Shen \inst{1}
          \and
          K. Rubinur \inst{1}
          \and
          A. Schimek \inst{1}
          \and
          A. Lasrado \inst{1}
          }
      \institute{\inst{1}Institute of Theoretical Astrophysics, University of Oslo, P.O. Box 1029, Blindern, 0315 Oslo, Norway\\ 
      \inst{2}Armagh Observatory and Planetarium, College Hill, Armagh BT61 9DG, UK \\ 
      \inst{3}Department of Physics and Astronomy, University College London, London, WC1E 6BT, UK \\ 
      \inst{4}INAF - Osservatorio Astronomico di Brera, Via Brera 28, 20121 Milano, Italy \\ 
      \inst{5}Dipartimento di Fisica e Astronomia Augusto Righi, Università degli Studi di Bologna, via Gobetti 93/2, 40129 Bologna, Italy\\ 
      \inst{6}INAF – Osservatorio di Astrofisica e Scienza dello Spazio di Bologna, Via Gobetti 101, 40129 Bologna, Italy \\ 
      \email{bendix.hagedorn@astro.uio.no}
		      }

   \date{Received 28 February 2024 / Accepted: 6 May 2024 }


\abstract{
We derived molecular gas fractions ($f_\mathrm{mol}=M_\mathrm{mol}/M_*$) and depletion times ($\tau_\mathrm{mol}= M_\mathrm{mol}/\mathrm{SFR} $) for 353 galaxies representative of the local star-forming population with $10^{8.5}\,M_\odot < M_* < 10^{10.5}\,M_\odot$ drawn from the ALLSMOG and xCOLDGASS surveys of CO(2-1) and CO(1-0) line emission. By adding constraints from low-mass galaxies and upper limits for CO non-detections, we find the median molecular gas fraction of the local star-forming population to be constant at $\log f_\mathrm{mol}=-0.99^{+0.22}_{-0.19}$, challenging previous reports of increased molecular gas fractions in low-mass galaxies. Above $M_*\sim 10^{10.5}\,M_\odot$, we find the $f_\mathrm{mol}$ versus $M_*$ relation to be sensitive to the selection criteria for star-forming galaxies. We tested the robustness of our results against different prescriptions for the CO-to-H$_2$ conversion factor and different selection criteria for star-forming galaxies. The depletion timescale $\tau_\mathrm{mol}$ weakly depends on $M_*$, following a power law with a best-fit slope of {\bx $0.16\pm 0.03$}. This suggests that small variations in specific star formation rate ($ \mathrm{sSFR=SFR}/M_*$) across the local main sequence of star-forming galaxies with $M_* < 10^{10.5}\,M_\odot$ are mainly driven by differences in the efficiency of converting the available molecular gas into stars. We tested these results against a possible dependence of $f_\mathrm{mol}$ and $\tau_\mathrm{mol}$ on the surrounding (group) environment of the targets by splitting them into centrals, satellites, and isolated galaxies, and find no significant variation between these populations. We conclude that the group environment is unlikely to have a large systematic effect on the molecular gas content of star-forming galaxies in the local Universe.
}

\maketitle
%
\section{Introduction}
The tight correlations found between the cold molecular gas content of galaxies and other physical galaxy properties, such as stellar mass ($M_*$) and star formation rate (SFR) are powerful and widely used tools in the study of galaxy evolution.
These quantities are inextricably linked because stars form in clouds of cold molecular gas.
However, the exact physical processes governing the interplay of cloud collapse and feedback are not fully understood and may vary among galaxy populations \citep[e.g.,][]{sunMolecularGasProperties2020}.
Scaling relations have the potential to reveal any systematic differences across regimes in $M_*$, SFR, and environment, but this requires data covering the associated parameter ranges.
Due to observational challenges, molecular line data in low-mass, low-metallicity galaxies are lacking.
As a consequence, scaling relations have so far been largely based on massive ($M_*>10^{10}\,\msun$), metal-rich galaxies (see \citealt{saintongeColdInterstellarMedium2022} for a review).
Recent studies have made efforts to push observations of cold molecular gas to lower masses with statistically significant sample sizes \citep{ciconeFinalDataRelease2017, saintongeXCOLDGASSComplete2017, wylezalekMASCOTESOARO2022}, but there is no clear picture of the behavior of molecular gas scaling relations in the low-mass regime ($M_*<10^{9}\,\msun$) yet.
Particularly in the case of intensive properties (i.e., those that do not directly depend on the size of the system), such as the molecular gas fraction ($\fmol=\mmol /M_*$), the results reported in the literature differ significantly.
Some recent studies find $\fmol$ to be anticorrelated with stellar mass following a simple power-law across the entire mass range studied \citep[e.g.,][]{huntScalingRelationsBaryonic2020}, while others report a shallow or flat relation at lower masses \citep[e.g.,][]{caletteHIH2TOSTELLARMASS2018, saintongeXCOLDGASSComplete2017} with a downturn above a certain stellar mass.
This feature is sometimes attributed to the larger fraction of passive galaxies at higher masses \citep{jiangSUBMILLIMETERTELESCOPECO2015}, but it appears even in samples selected to exclude passive galaxies \citep[e.g.,][]{saintongeXCOLDGASSComplete2017, caletteHIH2TOSTELLARMASS2018}.
The situation is similar for the molecular gas depletion time ($\tmol=\mmol/\mathrm{SFR}$), which is a measure of how rapidly galaxies convert their molecular gas into stars.
Some authors report that $\tmol$  is independent of $M_*$ for main-sequence galaxies \cite[e.g.,][]{accursoDerivingMultivariateACO2017, boselliColdGasProperties2014}, but others find $\tmol$ and $M_*$ to correlate, at least above $M_*=10^9\,\msun$ \citep[e.g.,][]{huntScalingRelationsBaryonic2020, saintongeXCOLDGASSComplete2017}.
A lack of relation between  $\tmol$ and $M_*$ would indicate that star formation proceeds at a universal rate in galaxies, while a correlation has been argued to be a consequence of global galaxy properties affecting processes on cloud scales \citep{sunMolecularGasProperties2020, saintongeColdInterstellarMedium2022}.
These differences hint at the uncertainties associated with measuring the molecular gas content of galaxies and using it to constrain scaling relations.
One such uncertainty is due to the fact that the H$_2$ molecule, which makes up the majority of the cold molecular gas, has no permanent dipole.
As a consequence, it is not readily excited to emit radiation at the densities and temperatures found in molecular clouds, providing no direct tracer for the majority of the gas.
Instead, emission arising from the J=1 to J=0 rotational transition of carbon monoxide (CO), the second most abundant molecule, is frequently used as a proxy for the molecular gas mass.
However, the abundance of CO in the molecular gas depends on the local metallicity and radiation field; the conversion factor between observed CO(1-0) luminosity and inferred molecular gas mass varies with the physical properties of the interstellar medium (ISM).
Many empirical and theoretical prescriptions for this conversion factor ($\alpha_\mathrm{CO}\equiv M_\mathrm{mol}/L^{\prime}_\mathrm{CO(1-0)}$) have been proposed \citep{accursoDerivingMultivariateACO2017,wolfireDARKMOLECULARGAS2010,gloverRelationshipMolecularHydrogen2011,bolattoCOtoHConversionFactor2013}, but there remains a large uncertainty in choosing a specific model, which goes some way to explain the different scaling relations reported in the literature.
In low-mass, low-metallicity galaxies, shielding is less effective so that very little CO can form, increasing the fraction of so-called ``dark'' molecular gas \citep{wolfireDARKMOLECULARGAS2010}.
This makes it even more difficult to obtain reliable estimates of the molecular gas mass of such galaxies, and studies targeting galaxies independently of their gas fraction generally have low CO  detection rates in this regime.
This leads to samples either including only the most gas-rich low-mass objects, or a large fraction of non-detections depending on the survey strategy.
The former has the potential to bias the sample toward a particular, non-representative class of objects, while the latter necessitates the use of analysis methods suitable to include upper limits derived from non-detections.
These differences in turn influence the scaling relations inferred from the data.
The environment in which galaxies evolve may present an additional confounding factor affecting their scaling relations.
Galaxies residing in large clusters have been found to adhere to different scaling relations than their isolated counterparts \citep[e.g.,][]{brownVERTICOVirgoEnvironment2021, zabelVERTICOIIHow2022}.
Most galaxies however reside in smaller groups, rather than the rare large clusters.
The effect of this less extreme environment is likely to be more subtle, with lower gas temperatures in the halo and fewer close interactions between members.
Nonetheless, examples of the group environment affecting a member galaxy have been found in both observations and simulations, raising the question whether the group environment systematically affects galaxy evolution \citep[e.g.,][]{leeALMAACACO2022, zhouImpactEnvironmentLives2022a}.
Tidal forces and feedback mechanisms can contribute to the transfer of gas from the ISM of the group members to the intra-group region while simultaneously affecting its physical properties, as observed in the Stephan's quintet \citep{appletonSHOCKENHANCEDEMISSIONDETECTION2013a, appletonMultiphaseGasInteractions2023a} and in the M81 triplet \citep{deblokHighresolutionMosaicNeutral2018}.
If this were the case in most galaxy groups, the environment would need to be accounted for as a factor in determining statistical scaling relations.
Single-dish CO surveys in the past decade have increased the available sample size to a level where it becomes feasible to study populations separately based on their environment, but this possibility has not yet been exploited to its full extent.
In this study, we present improved constraints on the scaling relations of cold molecular gas in local star-forming galaxies down to $M_*=10^{8.5}\,\msun$ based on a reanalysis of the APEX low-redshift legacy survey for molecular gas (ALLSMOG) \citep{ciconeFinalDataRelease2017} and extended CO Legacy Database for GASS (xCOLDGASS) \citep{saintongeXCOLDGASSComplete2017} surveys, which include rigorous upper limits for their non-detections.
Additional homogeneous SFR, $M_*$, metallicity, redshift, and environment data are available from catalogs compiled for the SDSS parent sample, from which both of these surveys were originally drawn.
\cite{ciconeFinalDataRelease2017} demonstrated that including the low-$M_*$ sample from ALLSMOG can significantly affect the slopes of scaling relations, but did not take the extra steps to convert CO luminosities into molecular gas masses, and so did not specifically delve into the $\fmol$ vs. $M_*$ or $\tmol$ vs. $M_*$ scaling relations that we aim to explore here.
In addition, we exploit the size of the sample and the availability of the necessary ancillary data products to investigate the impact of the group environment on molecular gas scaling relations which, to our knowledge, has never been done.
This is necessary groundwork to inform future studies that may want to control for the environment variable when investigating other effects, such as various modes of feedback, on the cold gas scaling relations.
This paper is structured as follows.
We describe the sample and its underlying selection criteria in Sect.~\ref{sec:sample} and the treatment of censored data in Sect.~\ref{sec:methods}.
The resulting scaling relations are presented and compared to previous studies of similar samples in Sect.~\ref{sec:scaling}.
In Sect.~\ref{sec:env} we explore the effect of the group environment.
Finally, we present our conclusions and discuss their implications in Sect.~\ref{sec:conclusion}.
Throughout this paper we assume a flat $\Lambda$CDM cosmology with $\Omega_\mathrm{m}=0.3$, $\Omega_\mathrm{\Lambda}=0.7$, and $H_0=70\,\mathrm{km\,s^{-1}\,Mpc^{-1}}$ for consistency with \citet{saintongeXCOLDGASSComplete2017} when deriving distance dependent quantities.


\section{Sample selection and methods}
\label{sec:sample}
\begin{figure*}[htb]
\centering
    \includegraphics[width=\columnwidth, clip=true, trim=0.25cm 0.7cm 0.5cm 1.cm]{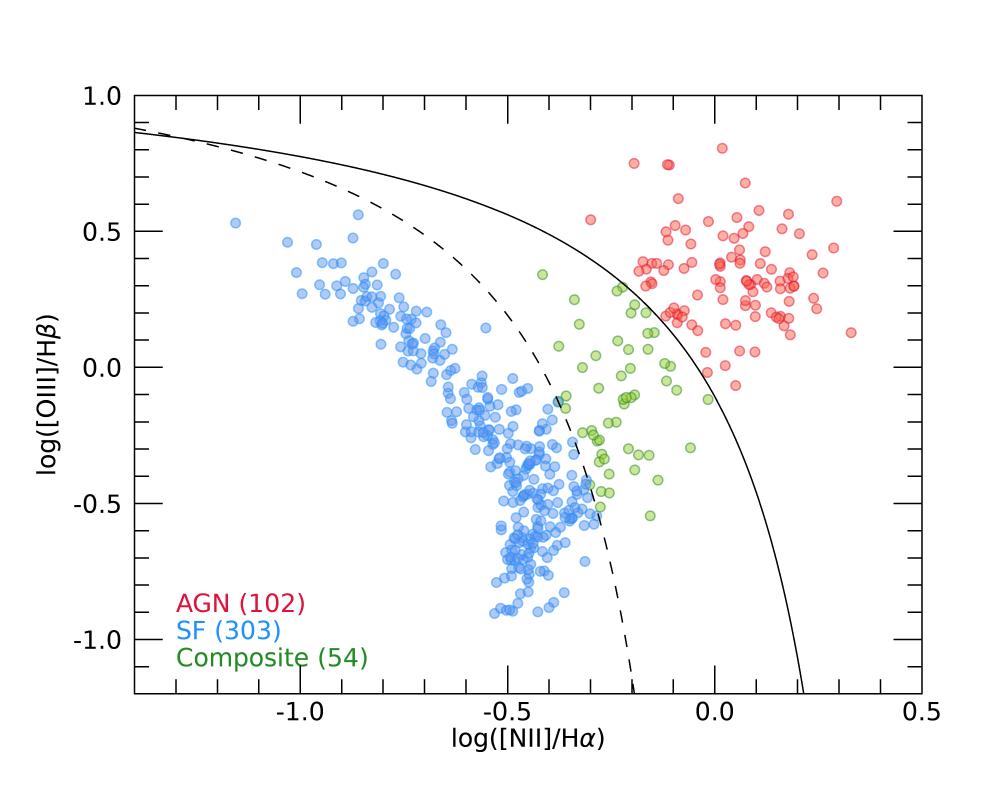}
    \includegraphics[width=\columnwidth, clip=true, trim=0.25cm 0.7cm 0.5cm 1.cm]{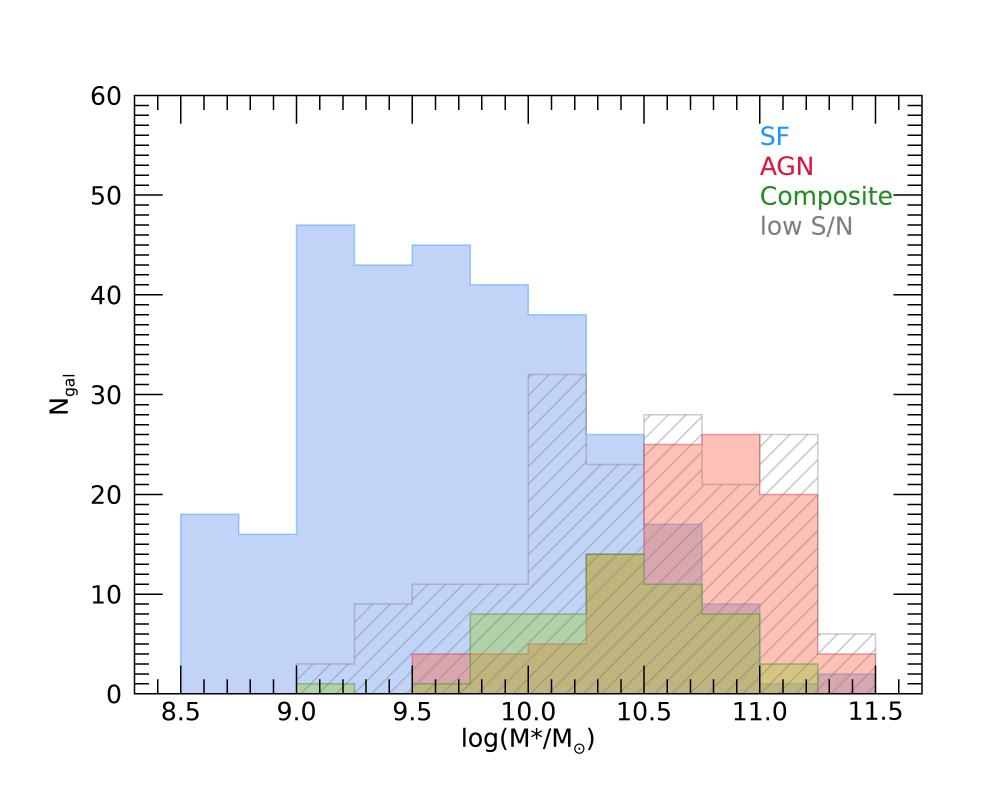}
     \caption{BPT classification fo the sample. \textit{Left:} BPT diagram of the full sample. The dashed line is the demarcation between pure SF galaxies and AGN or composite galaxies according to \citet{kauffmannHostGalaxiesActive2003}. The solid line is the theoretical limit for the ionization state produced by star formation according to \citet{kewleyTheoreticalModelingStarburst2001}, splitting the sample further into AGN and composite systems. \textit{Right:} Distribution of stellar masses in the SF, AGN, composite and low-S/N subsamples.}
   \label{fig:bpt}
\end{figure*}
We rely on observations of the CO(1-0) and CO(2-1) lines to infer the cold molecular gas properties used to constrain scaling relations.
From the available archival CO data we chose our sample to:
\begin{enumerate}
    \item be representative of the local star-forming galaxy population with $10^{8.5}\,\msun < M_* < 10^{10.5}\,\msun$;
    \item have available data on galaxy environment;
    \item have homogeneous ancillary data such as star formation rates (SFR) and stellar masses.
\end{enumerate}
The ALLSMOG \citep{ciconeFinalDataRelease2017} and xCOLDGASS \citep{saintongeXCOLDGASSComplete2017} surveys are drawn from the Sloan Digital Sky Survey (SDSS) DR7 \citep{SDSSDR72009} which ensures criteria (2) and (3) are met, while their survey strategies are well suited to achieve criterion (1), as is outlined in the following.
Section~\ref{sec:sf_selection} further describes how we select star-forming galaxies based on their star formation rates relative to the main sequence in the local universe.
Both surveys also target galaxies with available HI~21\,cm observations, which enables us to include the scaling relations of cool neutral hydrogen gas in our analysis.
\subsection{xCOLDGASS}
xCOLDGASS \citep{saintongeXCOLDGASSComplete2017} is comprised of 532 local galaxies with $M_*>10^{9}\,\msun$ and redshift $0.01<z<0.05$ observed in CO(1-0) with the IRAM 30-meter telescope.
Of these, 366 were observed in the original COLDGASS survey and have $M_*>10^{10}\,\msun$ and redshift $0.025<z<0.05$, while the remaining 166 were targeted in an effort to extend the survey into the range of $10^{9}\,\msun<M_*<10^{10}\,\msun$ and have $0.01<z<0.02$.
Targets were selected randomly from the SDSS parent sample based only on the mass and redshift criteria \citep[see][for details]{saintongeCOLDGASSIRAM2011}, resulting in a sample representative of local galaxies within the mass range.
xCOLDGASS did not specifically target star forming galaxies so we apply an additional selection criterion for this study, which we describe in Sect.~\ref{sec:bpt}.
\subsection{ALLSMOG}
ALLSMOG (\citealt{ciconeFinalDataRelease2017}, see also \citealt{bothwellALLSMOGAPEXLowredshift2014} for a previous incomplete release) contains a total of 97 local star-forming galaxies, of which 88 were observed in CO(2-1) with the APEX telescope and 9 more in CO(1-0) and CO(2-1) with the IRAM 30-meter telescope.
Their targets were selected from the SDSS in the same randomized manner as employed in xCOLDGASS with $10^{8.5}\,\msun<M_*<10^{10}\,\msun$, and $0.01<z<0.03$. 
Additionally, they select targets whose gas-phase metallicity, as initially reported in the MPA-JHU catalog computed according to the \citet{tremontiOriginMassMetallicity2004} calibration, is $\mathrm{12+\log(O/H)}>8.5$ for targets observed with APEX and $\mathrm{12+\log(O/H)}>8.3$ for observations with the more sensitive IRAM 30-meter telescope. 
However, in this work, metallicities are recalculated homogeneously for all of the sample for which this computation is possible (see details in Sect.~\ref{sec:bpt}) using the more recent calibration of the [OIII]-[NII] strong line index by \citet{curtiNewFullyEmpirical2017} (hereafter O3N2 C17), so the aforementioned limits do not directly translate.
Since the metallicity limit applied to ALLSMOG could potentially bias our sample,
we apply a weighting in our analysis to account for this effect, as discussed in more detail in Sect.~\ref{sec:methods}.
\subsection{BPT classification}
\label{sec:bpt}
In order to select the targets for which the gas-phase metallicity (needed to calculate the $\alpha_\mathrm{CO}$ conversion factor) can be directly computed using strong line calibration methods, we 
employ the Baldwin, Phillips \& Terlevich (BPT) diagram method \citep{baldwinClassificationParametersEmissionline1981}.
The BPT classification indicates the dominant excitation mechanism in the region covered by the optical spectra of the galaxies in the sample.
The left panel of Fig.~\ref{fig:bpt} shows the BPT diagram of the combined ALLSMOG and xCOLDGASS sample of 629 galaxies.
We calculated the emission line ratios reported in Fig.~\ref{fig:bpt} consistently for the entire sample using a full fit of the optical spectra performed with the \textsc{GandALF} code \citep{sarziSAURONProjectIntegralfield2006}, which fits stellar continuum and emission line templates simultaneously.
There are 170 galaxies with an amplitude-to-noise ratio $\mathrm{S/N}<3$ in at least one of the relevant emission lines (H$\alpha$, H$\beta$, [OIII], and [NII]) which are not classified in the BPT diagram, and so are not shown in the left panel of Fig.~\ref{fig:bpt}.
We designate these as ``low-S/N'' and show their stellar mass distribution in the right panel of Fig.~\ref{fig:bpt}.
The remaining galaxies with $\mathrm{S/N}>3$ in all lines are then classified based on their position in the BPT diagram, following
\citet{kauffmannHostGalaxiesActive2003} and \citet{kewleyTheoreticalModelingStarburst2001}. 
The sample is thus split into star forming (SF) galaxies, active galactic nuclei (AGN), and composite systems, where the SF subsample covers the entire range of $8.5<\log(M_*/\msun)<11.5$, while the AGN subsample has a much higher average stellar mass and only covers $9.5<\log(M_*/\msun)<11.5$, as shown in the right panel of Fig.~\ref{fig:bpt}.
The optical spectra used in this study sample a $3''$-diameter region, which is the size of the SDSS fiber.
Since our targets are low-redshift galaxies, they are quite extended in the sky and the fiber will only probe the very central region of the galaxy.
The BPT classification is therefore indicative of the dominant excitation mechanism in this central region only.
It is possible that we classify galaxies as AGN dominated or composite systems even though globally they may be mainly star forming.
However, if a galaxy is star forming in the central region, star formation is likely to be the dominant excitation mechanism globally as well, since AGN are generally located centrally.
Thus, we can be reasonably certain that galaxies classified as star-forming in the BPT diagram are not affected by current AGN activity.
This is important for the metallicity calculations outlined in Sect.~\ref{sec:alpha_co_method}, which are based on emission line ratios that may be affected by the presence of an AGN. 
Indeed, the purpose of the BPT classification discussed here is only to select the galaxies for which we can reliably apply strong line calibration recipes to compute the gas phase metallicity.
\subsection{``Star-forming'' classification}
\label{sec:sf_selection}
\begin{figure}[htb]
\centering
    \includegraphics[clip=true, trim=0.25cm 0cm 0.1cm 0.1cm, width=\columnwidth]{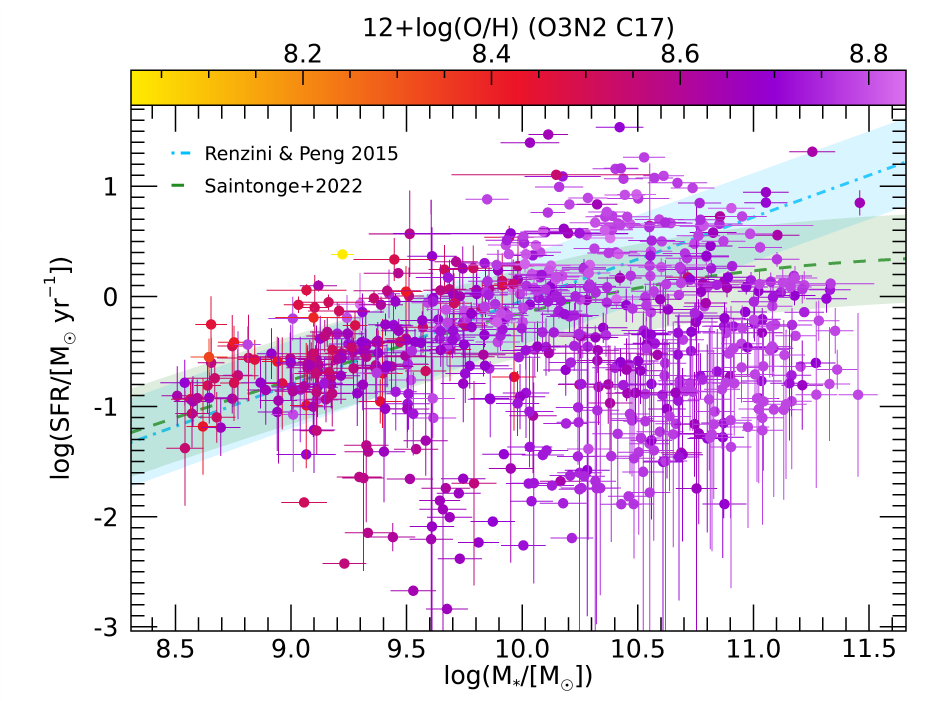}
     \caption{SFR vs. stellar mass for the 625 ALLSMOG + xCOLDGASS galaxies that have a reliable SFR in the MPA-JHU catalog. The color scale corresponds to the gas phase metallicity, which was calculated from the O3N2 line index \citep{curtiNewFullyEmpirical2017} for galaxies classified as SF or composite in the BPT diagram and according to the mass-metallicity relation \citep{curtiMassMetallicityFundamental2020} for those classified AGN and low-S/N. The blue dash-dotted line shows the empirical relation for the star forming main sequence found by \citet{renziniOBJECTIVEDEFINITIONMAIN2015} including a 0.4\,dex scatter (blue shaded). The green dashed line shows the star forming main sequence following \citet{saintongeColdInterstellarMedium2022}.}
   \label{fig:mstar_sfr}
\end{figure}

Figure~\ref{fig:mstar_sfr} shows the SFR of the 625 ALLSMOG and xCOLDGASS galaxies for which we have a reliable SFR measurement plotted against their stellar mass.
Both quantities are obtained from the MPA-JHU catalog\footnote{\url{https://live-sdss4org-dr12.pantheonsite.io/spectro/galaxy_mpajhu/}} for SDSS DR8, thus ensuring a homogeneous methodology across the entire sample.
Also shown are two empirical relations for the star-forming main sequence of galaxies (SFMS) reported by \citet{renziniOBJECTIVEDEFINITIONMAIN2015} and \citet{saintongeColdInterstellarMedium2022} with an added 0.4\,dex scatter.

As we aim to study specifically star-forming galaxies in the local universe, we need to apply an additional selection to this sample.
Applying a uniform selection across the entire sample also ensures that we homogenize the ALLSMOG (which specifically targeted star-forming galaxies) and xCOLDGASS (which did not) data.
We chose a criterion based on the offset from the main sequence of star-forming galaxies following the definition of \citet{renziniOBJECTIVEDEFINITIONMAIN2015}.
We calculate the offset from the main sequence as 
\begin{equation}
\Delta(\mathrm{MS}) = \mathrm{sSFR_{measured}/sSFR_{MS}},
\end{equation}
where $\mathrm{sSFR_{measured}}=\mathrm{SFR_{measured}}/M_{*,\mathrm{measured}}$ is the measured specific star formation rate and $\mathrm{sSFR_{MS}}$ is the specific star formation rate predicted for the object by the \citet{renziniOBJECTIVEDEFINITIONMAIN2015} relation based on its stellar mass.
In our analysis of molecular gas scaling relations, we exclude galaxies with $\log\Delta(\mathrm{MS})<-0.4$\,dex, to avoid contaminating our sample with quenched passive galaxies.
For the remainder of this work we refer to the sample selected as outlined above simply as the ``star-forming sample''.
When using a sample based on different selection criteria this is stated explicitly.
The star-forming sample contains 353 galaxies, 283 of which are detected in at least one of the CO(1-0) and CO(2-1) emission lines and 70 of which are non-detections in CO.
Starburst galaxies, which lie above the main sequence in the $M*$-SFR plane, are included in this selection.
The possible additional dependence of $\alpha_\mathrm{CO}$ on the dynamic state of the gas is properly taken into account when calculating molecular gas masses for the starburst galaxies in our sample, as described in Sect.~\ref{sec:alpha_co_method}.
There are a number of different approaches for selecting star-forming galaxies, which will generally result in different slopes and shapes of the main sequence relation (see \citealt{pearsonInfluenceStarformingGalaxy2023} for an overview).
In Fig.~\ref{fig:mstar_sfr} we show both a linear main sequence relations \citep{renziniOBJECTIVEDEFINITIONMAIN2015} and one with a high-mass turnover \citep{saintongeColdInterstellarMedium2022}.
This illustrates that while the composition of our final sample will depend on the main sequence definition we adopt here, the key differences arise in the high-mass regime ($M>10^{10.5}\,\msun$).
The low-mass regime, on which we focus in our analysis, is unlikely to be affected significantly by this choice.
We test this assumption by comparing between the results obtained for the \citet{renziniOBJECTIVEDEFINITIONMAIN2015} and \citet{saintongeColdInterstellarMedium2022} main sequence samples respectively for key scaling relations later on.
We further explore how our choice of selection criteria affects our results by comparing it to other common approaches in Appendix \ref{app:sf_criteria}.
\subsection{Metallicity and CO-to-H$_2$ conversion factor}
\label{sec:alpha_co_method}
Obtaining a galaxy's molecular gas mass from its integrated CO(1-0) line luminosity is not straightforward.
Most of the molecular gas mass is contained in H$_2$, which we do not observe directly, while CO only makes up a small fraction of the total mass.
More importantly, the abundance of CO depends on the local gas chemistry and radiation field, so that a conversion factor obtained from a local cloud in the Milky Way need not apply to another galaxy.
The main parameter governing CO abundance is thought to be the gas-phase metallicity, with both observations and theoretical predictions pointing at a lower CO abundance in low-metallicity galaxies \citep[e.g][]{wolfireDARKMOLECULARGAS2010, maddenTracingTotalMolecular2020}.
The ratio of CO to H$_2$ gas changes with metallicity in this picture, necessitating a metallicity dependent $\alpha_\mathrm{CO}$ conversion factor.
\begin{figure*}[htb]
\centering
    \includegraphics[width=\textwidth, clip=true, trim=0cm 0.1cm 0.5cm 0.7cm]{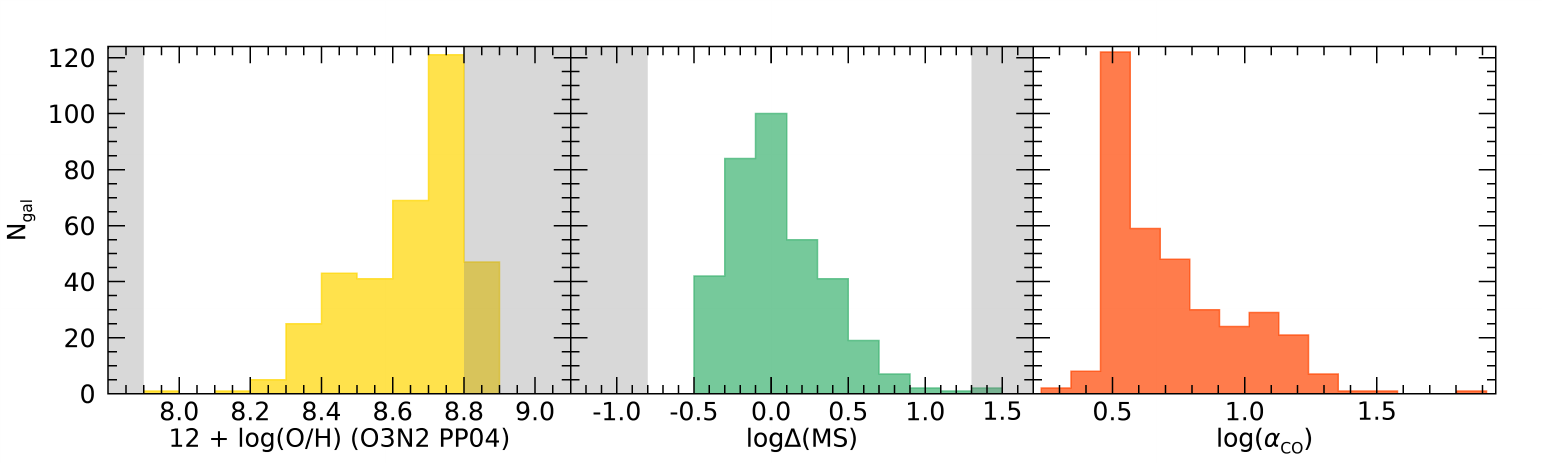}
     \caption{Distribution of our sample in gas phase metallicity (\textit{left}), distance off the main sequence (\textit{center}), and $\alpha_\mathrm{CO}$ (\textit{right}). Shaded areas indicate the parameter space where \citet{accursoDerivingMultivariateACO2017} suggest adapting their prescription for $\alpha_\mathrm{CO}$.}
   \label{fig:aco_dist}
\end{figure*}

The color scale in Fig.~\ref{fig:mstar_sfr} corresponds to the gas-phase metallicity of our sample in units of 12+log(O/H), which, for galaxies that were classified as star-forming or composite in the BPT diagram, was calculated from measured fluxes of the Balmer $\alpha$ and $\beta$ lines and the [NII]$\lambda 6583$ and [OIII]$\lambda 5007$ forbidden lines following \citet{curtiNewFullyEmpirical2017}.
Galaxies classified as AGN instead have their metallicity assigned according to the prediction from the mass-metallicity-relation as defined by \citet{curtiMassMetallicityFundamental2020}, because the strong-line diagnostics presented in \citet{curtiNewFullyEmpirical2017} are only calibrated for star-forming galaxies.
Most AGN in our sample lie above $M_*=10^{10}\,\msun$, where the mass-metallicity relation flattens, so that we do not expect our assumptions on their metallicity to add significant additional uncertainty to our molecular gas scaling relations.
To determine the $\alpha_\mathrm{CO}$ value we use the prescription presented in \citet{accursoDerivingMultivariateACO2017}: 
\begin{equation}
\label{eq:alphaco}
\begin{split}
\log\alpha_\mathrm{CO} = \, & 14.752 - 1.623\cdot [12 + \log\mathrm{(O/H)}] \\
& + 0.062\cdot\log\Delta(\mathrm{MS}).
\end{split}
\end{equation}
{\bx Here, the gas phase metallicity in units of 12 + log(O/H) is calculated according to the calibration by \citet{pettiniIiiIiAbundance2004} using the same line fluxes as outlined above (hereafter O3N2 PP04), in order to be fully consistent with \citet{accursoDerivingMultivariateACO2017}.
For AGN, we use the mass-metallicity relation defined by \cite{kewleyMetallicityCalibrationsMass2008}, which uses the PP04 calibration.}
Equation \ref{eq:alphaco} contains a secondary dependence on the local radiation field through the parameter $\Delta(\mathrm{MS})=\mathrm{SFR_{measured}}/\mathrm{SFR_{MS}}$, the distance in SFR off the star forming main sequence (SFMS).
In this instance, $\mathrm{SFR_{MS}}$ corresponds to the redshift dependent analytic definition of the main sequence by \citet{whitakerSTARFORMATIONMASS2012}.
Both metallicity and $\log\Delta(\mathrm{MS})$ measurements come with associated uncertainties, that we propagate to estimate the uncertainty in the resulting conversion factor.
The median uncertainty on the metallicity values, derived from the uncertainties of the line ratio measurements, is {\bx 0.025 in units of 12+log(O/H) (O3N2 PP04).}
For $\log\Delta(\mathrm{MS})$ the median error is 0.16\,dex.
\citet{accursoDerivingMultivariateACO2017} give the parameter ranges in which their empirical $\alpha_\mathrm{CO}$ is constrained as $7.9 < 12 + \log\mathrm{(O/H)} < 8.8$ in metallicity (O3N2 PP04) and $-0.8 < \log\Delta(\mathrm{MS}) < 1.3$
Our sample spans {\bx $ 7.94 < 12 + \log\mathrm{(O/H)} < 8.88$  (O3N2 PP04)} and $-0.49 < \log\Delta(\mathrm{MS}) < 1.32$, so the majority of our galaxies lie in the well constrained regime.
The full distribution of our star-forming sample in metallicity, $\log\Delta(\mathrm{MS})$, and the resulting $\alpha_\mathrm{CO}$ is shown in Fig.~\ref{fig:aco_dist}.
For those galaxies with $12 + \log\mathrm{(O/H)} > 8.8$ we follow the suggestions of \citet{accursoDerivingMultivariateACO2017} and calculate $\alpha_\mathrm{CO}$ using $12 + \log\mathrm{(O/H)} = 8.8$.
For starburst galaxies with $\log\Delta(\mathrm{MS}) > 1.3$, \citet{accursoDerivingMultivariateACO2017} recommend a lower value of $\alpha_\mathrm{CO}$, because dynamical effects dominate over metallicity in setting the $\alpha_\mathrm{CO}$ in this regime.
We adopt a value of $\alpha_\mathrm{CO} = (1.7\pm0.5)\,\acounits$ based on a recent study of local ultra luminous infrared galaxies (ULIRGs) by \citet{montoyaarroyaveSensitiveAPEXALMA2023}, which is consistent with recent estimates of $\alpha_\mathrm{CO}$ values in other local starbursts.
This lower value is used to account for the effect of turbulence and high velocity dispersion in starbursts lowering the optical depth of the CO emission and so boosting the luminosity of CO line emission.
The conversion factors thus obtained for the galaxies in our star-forming sample range from $1.7\,\acounits$ {\bx ($2.75\,\acounits$ for non starbursts) to $82.2\,\acounits$} (see Fig.~\ref{fig:aco_dist}).
All values quoted here include a factor of $1.36$ to account for the presence of helium in the molecular gas, based on the helium mass fraction measured in the Milky Way \citep{asplundChemicalCompositionSun2009}.
Not all galaxies in our sample are observed in the CO(1-0) line.
For the ALLSMOG galaxies with only CO(2-1) observations, we assume $r_{21}\equiv L^{\prime}_\mathrm{CO(2-1)}/L^{\prime}_\mathrm{CO(1-0)}=0.7\pm0.2$, following \citet{leroyLowJCOLine2022} to obtain CO(1-0) luminosities.
We use aperture corrected values for all CO(1-0) and CO(2-1) luminosities following the corrections applied in the original ALLSMOG \citep{ciconeFinalDataRelease2017} and xCOLDGASS \citep{saintongeXCOLDGASSComplete2017} surveys.
Finally, we calculate the molecular gas masses for our sample as
\begin{equation}
    M_\mathrm{mol} = \alpha_\mathrm{CO}(12 + \log\mathrm{(O/H)},\Delta(\mathrm{MS}))\cdot L'_\mathrm{CO},
\end{equation}
where $L'_\mathrm{CO}$ is the aperture corrected CO(1-0) line luminosity and we are using the $\alpha_\mathrm{CO}$ prescription from Eq. \ref{eq:alphaco}.
To ensure the robustness of our results regarding the chosen prescription for $\alpha_\mathrm{CO}$, we test two additional models.
The results are described in Appendix \ref{app:alphaco}.
\subsection{Environment classification}
\label{sec:env_methods}
We use a galaxy group catalog\footnote{\url{https://www.galaxygroupfinder.net/catalogs}} by \citet{tinkerSelfCalibratingHaloBasedGroup2021} to classify the environment of each galaxy in our sample.
The catalog is constructed using a halo-based group finder algorithm which assigns galaxies to halos based on halo mass and distance in redshift space.
Galaxies are classified either as satellites or centrals in the catalog based on these criteria.
For a more detailed description of the algorithm see \citet{tinkerSelfCalibratingHaloBasedGroup2021} and references therein.
We add a third class, isolated galaxies, which are assigned to a group in which they are the only member.
In our sample of 353 star-forming local galaxies we identify 38 centrals, 42 satellites, and 247 isolated galaxies.
There are also 26 galaxies which are not contained in the group catalog, because it is not based on the original SDSS data but on the NYU value added galaxy catalog \citep{blantonNewYorkUniversity2005}.
The groups in our sample have halo masses of $11.0 < \log (M_\mathrm{h}/\msun) < 14.9$.
Galaxy groups are usually defined by a halo mass of $13 < \log(M_\mathrm{h}/\msun) < 14$ as opposed to clusters  with $\log(M_\mathrm{h}/\msun) > 14$.
In our sample only 5 galaxies reside in halos with $\log(M_\mathrm{h}/\msun) > 14$, so we probe mainly group environments and isolated objects.

\begin{figure*}[htb]
\centering
	\includegraphics[clip=true, trim=0.25cm 0cm 0.1cm 0.1cm, width=\columnwidth]{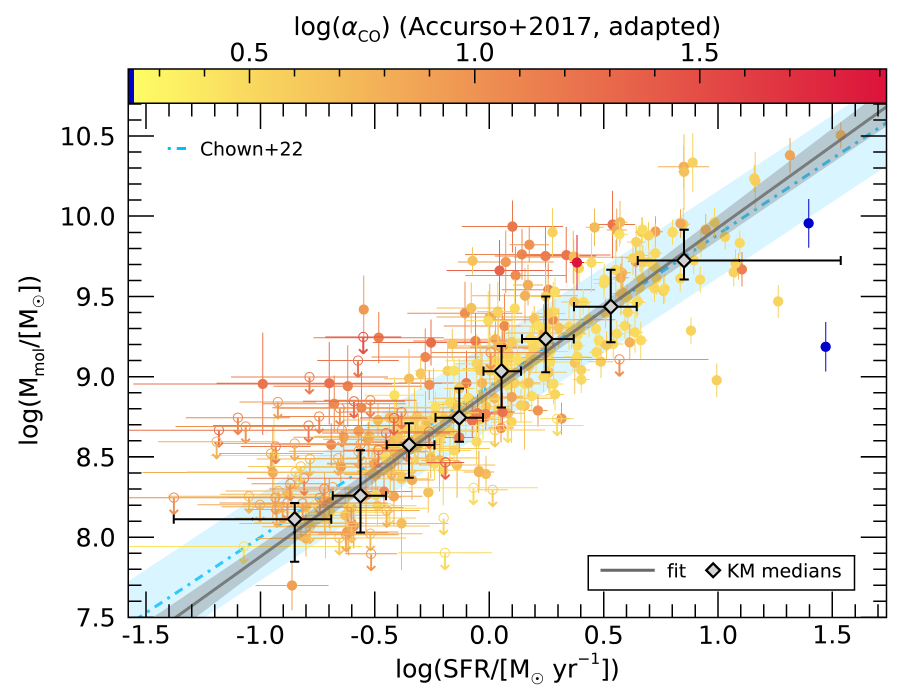}
    \includegraphics[clip=true, trim=0.25cm 0cm 0.1cm 0.1cm, width=\columnwidth]{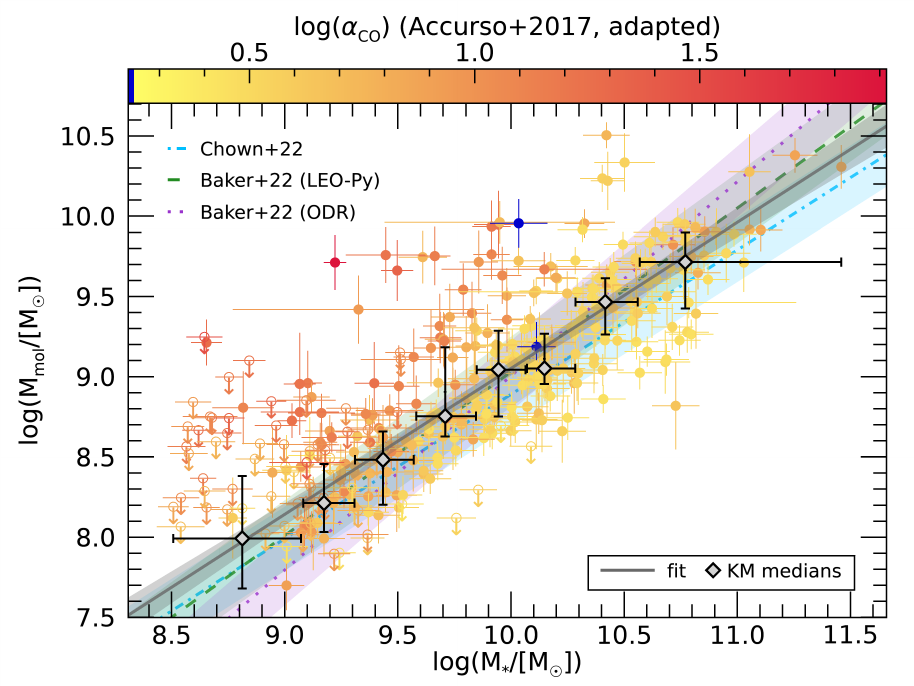}
    \caption{Scaling relations of the molecular gas mass. \textit{Left:} Molecular gas mass vs. SFR for the star-forming sample. For comparison we show the integrated inverse Schmidt-Kennicutt relation (blue dash-dotted) and corresponding intrinsic scatter reported by \citet{chownColdGasDust2022}. 
    \textit{Right}: Molecular gas mass vs. $M_*$, with the best-fit MGMS from \citet{chownColdGasDust2022} (blue dash-dotted) and \citet{bakerMolecularGasMain2022} with only detections (purple dotted) and detections and upper limits (green dashed). The color scale corresponds to the adopted CO-to-H$_2$ conversion factor.
    In both panels the solid gray line shows the best fit from the Bayesian linear regression {\bx to the unbinned data} with the shaded area indicating the 95\% confidence band. The black diamonds correspond to the binned medians computed using the Kaplan-Meier estimator. {\bx The associated error bars show the extent of the bin (in $x$) and the 25-75 percentile interval (in $y$)}. Open symbols with downward arrows denote upper limits. The navy area of the color bar indicates the specific value of the conversion factor used for extreme starbursts as described in Sect.~\ref{sec:alpha_co_method}.}
    \label{fig:sfr_mmol}
\end{figure*}
\section{Statistical methods}
\label{sec:methods}
Scaling relations are frequently expressed as power-law functions of the form
\begin{equation}
    Y = AX^b,
\end{equation}
making it convenient to fit the data in log-log space where the power-law is transformed into a linear function with slope $b$ and intercept $\log(A)$.
To fit this model to our data we use a Bayesian linear regression method implemented in the \textsc{IDL} procedure \textsc{linmix\_err} \citep{kellyAspectsMeasurementError2007}.
This procedure is capable of accounting for measurement errors in both the independent and dependent variables as well as left-censored data (i.e., upper limits) in the independent variable, making it ideally suited for our purposes.
To account for possible biases due to the somewhat skewed distribution of our sample in metallicity at low masses (see Sect.~\ref{sec:sample}), we apply a weighting to our fitting procedure.
To obtain the appropriate weights we compare the metallicity distribution of our sample to that of the sample of all star-forming galaxies selected from the MPA-JHU catalog (using the same criteria we applied to our sample).
The weights are calculated for bins of 0.2\,dex in stellar mass and 0.1\,dex in gas-phase metallicity.
It should be noted that this method is somewhat limited by the range of metallicities in our sample, since we cannot use weights to account for the lowest metallicities, which our sample does not cover.
However, our sample covers a range of metallicities that contains at least 90\%+ of galaxies in the MPA-JHU sample in each mass bin, giving us reasonable certainty that the weighting procedure can account for most of the potential bias in metallicity.
We discuss this weighting procedure in more detail in Appendix~\ref{app:met_weights}.
The \textsc{linmix\_err} procedure calculates the linear correlation coefficient $\rho_\mathrm{corr}$ between the variables in addition to returning a best-fit linear function.
As an additional measure for statistical independence in our data we adopt an estimator for Kendall's tau statistic proposed by \citet{oakesConsistencyKendallTau2008} that takes into account the presence of upper limits.
To compute this estimator we analyze pairs of bivariate data $(X_i,Y_i)$ and $(X_j,Y_j)$, where $i\neq j$.
Such a pair is said to be concordant if $(X_i-X_j)\cdot(Y_i-Y_j)>0$ and discordant if $(X_i-X_j)\cdot(Y_i-Y_j)<0$.
The estimator is then computed as
\begin{equation}
    \tau_\mathrm{Kendall} = \frac{C-D}{C+D},
\end{equation}
where $C$ is the number of definite concordances among the $n(n-1)/2$ pairs of data and $D$ is the number of definite discordances.
Due to the presence of upper limits, some pairs cannot be determined to be discordant or concordant.
This fact is taken into account by renormalizing the original Kendall's tau (for details see \citealt{oakesConsistencyKendallTau2008}).
The resulting measure can range in value from -1 to 1 with values close to zero indicating statistical independence of the two variables.
To test for nonlinear trends, we additionally bin the data and employ survival analysis in the form of the nonparametric Kaplan-Meier estimator \citep{KaplanMeier1958} to calculate the median for each bin.
{\bx We follow the method outlined in \citet{LeeWangStatisticalMethods}, which is well suited to retrieve a median even for bins with detection fractions as low as 20\% \citep[e.g.,][]{caletteHIH2TOSTELLARMASS2018}, albeit with large uncertainties.}
Consequently, we are able to estimate median cold gas masses and derived quantities even for the low-mass galaxies, where our data are dominated by CO non-detections.
We choose bin sizes such that each bin is populated by roughly the same number of galaxies to ensure that the calculated median is not dominated by random fluctuations due to low number statistics.
As a consequence, the bins at both ends of the mass distribution (Fig.~\ref{fig:bpt}) are considerably larger than at the centre.
%

\section{Gas scaling relations}
\label{sec:scaling}
Having established that our sample of 353 galaxies is representative of the local star-forming population with $10^{8.5}\,\msun < M_* < 10^{10.5}\,\msun$, we use it to derive scaling relations of $\mmol$, $\fmol$, and $\tmol$ for this population.
We apply the methods described in Sect.~\ref{sec:methods} to ensure that non-detections are accounted for in our scaling relations and analyze their impact by comparing the results to the literature.
\begin{table*}[htb]
    \caption{Results from Bayesian linear regression.}
    \centering
    \begin{tabular}{llccccc}
    \hline
    \hline
    $x$ & $y$ & intercept & slope & $\sigma$\,[dex] & $\rho_\mathrm{corr}$ & $\tau_\mathrm{Kendall}$\\
    \hline
    $\log (M_*/[\msun])$ & $\log (M_\mathrm{mol}/[\msun])$ & $-0.04\pm  0.38$ & $\hphantom{-}0.91\pm  0.04$ & $ 0.36$ & $\hphantom{-}0.84$ & $\hphantom{-}0.66$\\
    $\log (\mathrm{SFR}/[\msun\,\mathrm{yr^{-1}}])$ & $\log (M_\mathrm{mol}/[\msun])$ & $\hphantom{-}8.90\pm  0.02$ & $\hphantom{-}1.02\pm  0.04$ & $ 0.29$ & $\hphantom{-}0.89$ & $\hphantom{-}0.71$\\
    $\log (M_*/[\msun])$ & $\log f_\mathrm{mol}$ & $-0.38\pm  0.38$ & $-0.11\pm  0.04$ & $ 0.36$ & $-0.19$ & $-0.11$\\
    $\log (M_*/[\msun])$ & $\log(\tau_\mathrm{mol}/[\mathrm{yr}])$ & $\hphantom{-}7.29\pm  0.32$ & $\hphantom{-}0.16\pm  0.03$ & $ 0.28$ & $\hphantom{-}0.33$ & $\hphantom{-}0.21$\\
    $\log (M_*/[\msun])$ & $\log f_\mathrm{HI}$ & $\hphantom{-}6.16\pm  0.36$ & $-0.65\pm  0.04$ & $ 0.38$ & $-0.73$ & $-0.54$\\
   \hline
    \end{tabular}
    \caption*{Notes: Values quoted are best-fit intercept and slope with $1\sigma$ uncertainties from the Bayesian linear fit to the molecular and neutral gas scaling relations of the star-forming sample. Also listed are the intrinsic scatter $\sigma$ and the linear correlation coefficient $\rho_\mathrm{corr}$ returned by the fitting procedure. The final column contains an estimate for Kendall's $\tau$ correlation measure obtained taking into account the presence of upper limits.}
    \label{tab:fit_params}
\end{table*}
\subsection{Molecular gas mass}
\label{sec:fmol_results}
We begin by investigating the relationships between the molecular gas mass, the star formation rate and the stellar mass.
The former scaling relation is known as the integrated Kennicutt-Schmidt (KS) law \citep{schmidtRATESTARFORMATION, kennicuttStarFormationLaw1989} while the latter is often referred to as the molecular gas main sequence (MGMS) and both are frequently used in galaxy evolution studies, allowing us to compare our results with previous work \citep[e.g.,][]{chownColdGasDust2022, bakerMolecularGasMain2022}.
The molecular gas mass in our sample is strongly correlated with SFR and $M_*$ as can be seen in Fig.~\ref{fig:sfr_mmol}.
This is expected, since all these properties are extensive (i.e., they scale with overall galaxy mass).
More massive galaxies will typically contain larger amounts of cold gas which leads to the formation of more stars.
While more massive galaxies are also more often passive quenched systems, with smaller cold gas reservoirs, we exclude passive galaxies from our sample explicitly to remove their effect on the scaling relations.
For both the integrated SK-relation and the MGMS, the best-fit slope for our sample from the Bayesian linear regression is close to unity (see Table~\ref{tab:fit_params} and Fig.~\ref{fig:sfr_mmol}).
We note that the integrated SK relation is more commonly reported with $\mmol$ on the $x$ and SFR on the $y$-axis, but we chose the assignment of the axes according to how the variables are treated in the fitting procedure.
For comparison, we show empirical integrated SK and MGMS relations from recent studies \citep{chownColdGasDust2022, bakerMolecularGasMain2022} in Fig.~\ref{fig:sfr_mmol}.
This illustrates that our results are consistent with previous works.
However, it should be noted that both \citet{chownColdGasDust2022} and \citet{bakerMolecularGasMain2022} use a constant CO-to-H$_2$ conversion factor and their samples consist of galaxies with $\log (M_*/\msun) > 9$.
We test how the assumption of a constant as opposed to a metallicity dependent conversion factor affects our results in Appendix \ref{app:alphaco}.
\begin{figure*}[htb]
\centering
	\includegraphics[clip=true, trim=0.25cm 0cm 0.1cm 0.1cm, width=0.8\textwidth]{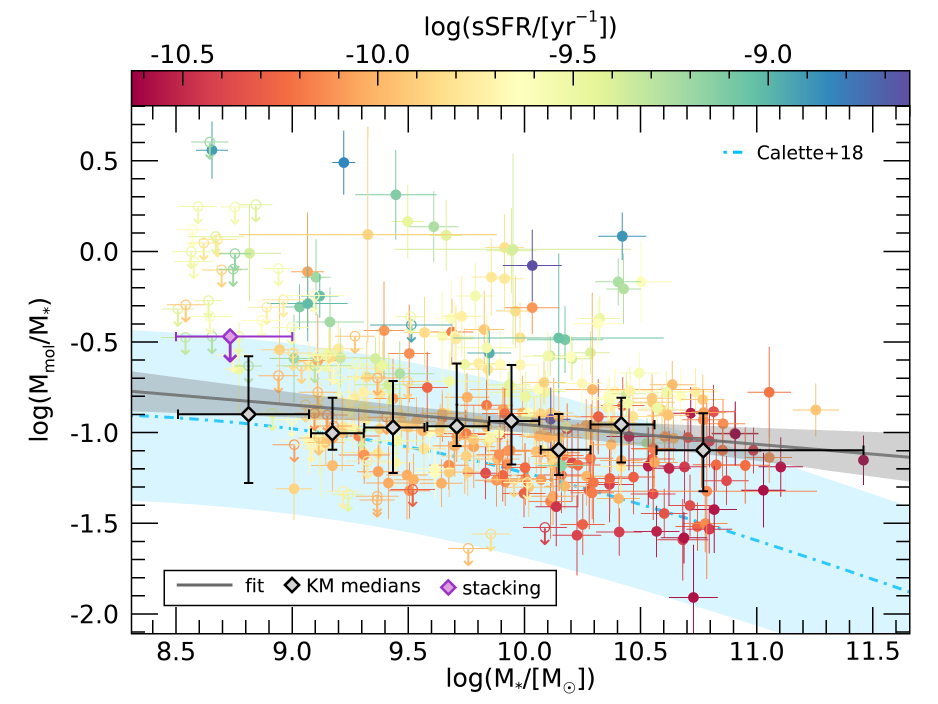}
     \caption{Molecular gas mass fraction vs. stellar mass for the star-forming sample. The solid gray line shows the best fit from the Bayesian linear regression {\bx to the unbinned data} with the shaded area indicating the 95\% confidence band. The black diamonds correspond to the binned medians computed using the Kaplan-Meier estimator. {\bx The associated error bars show the extent of the bin (in $x$) and the 25-75 percentile interval (in $y$)}. Open symbols with downward arrows denote upper limits. {\bxx The purple diamond shows the upper limit derived from spectral stacking of the 22 ALLSMOG CO(2-1) non-detections with masses below $M_*=10^9\,\msun$.} The blue dash-dotted line shows the scaling relations reported by \citet{caletteHIH2TOSTELLARMASS2018} based on their sample of late-type galaxies.}
   \label{fig:mstar_fmol}
\end{figure*}

\subsection{Molecular gas fraction}
Given the expected strong scaling between extensive galaxy properties in our sample, the next step is to look at the corresponding intensive properties such as the gas fraction ($\fmol=M_\mathrm{mol}/M_*$).
Dividing by stellar mass removes the primary scaling and provides insight into the physical processes behind the formation of cold gas and stars, and their dependence on galaxy stellar mass.
In Fig.~\ref{fig:mstar_fmol}, we plot $\fmol$ against $M_*$ for our sample and find a flat trend with no significant correlation as quantified by the linear correlation coefficient from the fitting procedure and the adopted estimator for Kendall's rank correlation coefficient (see Table~\ref{tab:fit_params}).
The binned medians shown in Fig.~\ref{fig:mstar_fmol} are also consistent with the fitted constant molecular gas fraction; The 25-75 percentile intervals derived from survival analysis overlap with the linear regression for all bins.
While the Kaplan-Meier estimator is capable of retrieving a median even for the lowest mass bin, which is dominated by non-detections, the corresponding uncertainties become large.
{\bxx We tested the validity of the KM medians for the low-mass regime by stacking the spectra of the 22 ALLSMOG non-detections of CO(2-1) with masses below $M_*=10^9\,\msun$ contained in our sample.
Stacking did not resulted in a clear detection, but we computed an upper limit on the typical molecular gas fraction in this mass range using the same procedure as for individual sources following \citet{ciconeFinalDataRelease2017}.
This upper limits is shown as a purple diamond in \ref{fig:mstar_fmol} and is consistent with the KM median value for the lowest mass bin. 
The stacked spectrum and the procedure used to obtained it are shown in \ref{app:apex_stacks}.
We note that since the stacking produces only an upper limit it does not rule out a downturn of the $\fmol$ versus $M_*$ relation below $M_*\sim 10^9\,\msun$.}
{\bx We have further tested the robustness of the KM medians against different binning choices (specifically, using 4, 5 and 6 bins instead of 9) and we confirm that each binning size delivers the same flat trend as seen here.
The bin size shown here is the one that maximizes the number of bins while still including enough objects in each bin to enable the survival analysis.}
We apply the same survival analysis method to the entire star-forming sample to determine the median gas fraction around which the local star-forming population is distributed, and find a median value of {\bx $\log\fmol=-0.99^{+0.22}_{-0.19}$ ($\log\fmol=-1.01^{+0.24}_{-0.23}$ when considering only galaxies with $M_* < 10^{10.5}\,\msun)$. Here the quoted errors correspond to the 25th and 75th percentiles.}
\begin{figure}[htb]
\centering
	\includegraphics[clip=true, trim=0.25cm 0cm 0.1cm 0.1cm, width=\columnwidth]{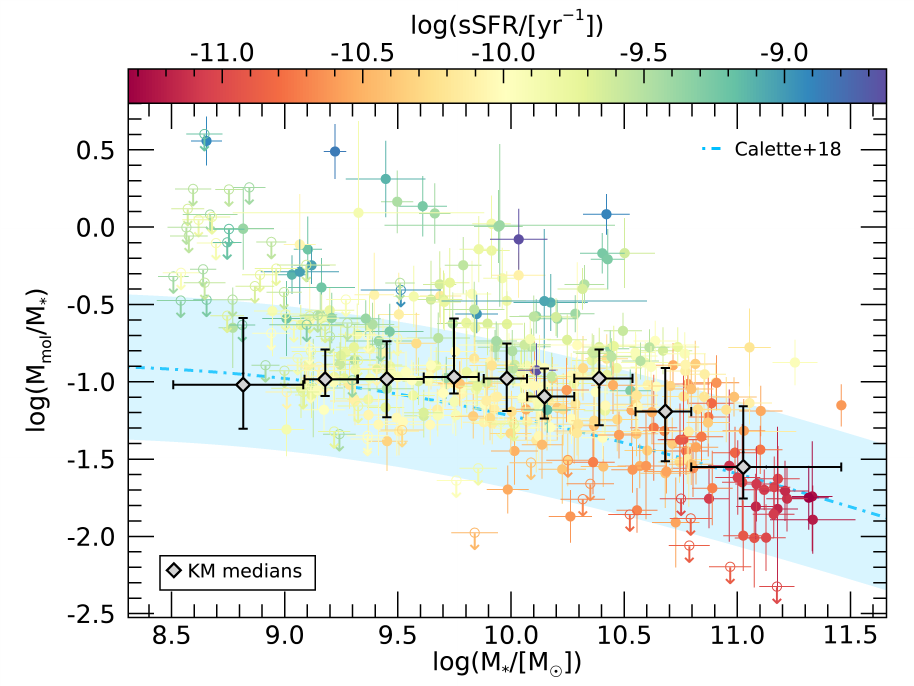}
     \caption{Same scaling relation as in \ref{fig:mstar_fmol} but for the star-forming plus starburst sample defined using the main sequence definition from \citet{saintongeColdInterstellarMedium2022}.}
   \label{fig:mstar_fmol_altMS}
\end{figure}

A flat or nearly flat relation between $\fmol$ and $M_*$ in the stellar mass range $10^{9}\,\msun < M_* < 10^{10}\,\msun$ is well supported by previous findings \citep[e.g.,][]{saintongeXCOLDGASSComplete2017, caletteHIH2TOSTELLARMASS2018}.
The picture is somewhat less clear at high masses ($M_*\gtrapprox 10^{10}\,\msun$), where reported scaling relations are sensitive to sample selection, and in the very low-mass regime ($M_*<10^{9}\,\msun$).
For the total xCOLDGASS sample, \citet{saintongeXCOLDGASSComplete2017} reported a mild negative trend between $\fmol$ and $M_*$ for main sequence galaxies with $10^{9}\,\msun<M_*<10^{10.5}\,\msun$ and a sharp drop-off at higher masses.
The fact that we do not observe this drop off is a direct consequence of how our star-forming sample is selected.
\citet{saintongeXCOLDGASSComplete2017} adopt a criterion based on the distance from the main sequence as defined in \citealt{saintongeMolecularAtomicGas2016} and updated in \citealt{saintongeColdInterstellarMedium2022} which flattens at high masses, as opposed to the linear main sequence derived in \citet{renziniOBJECTIVEDEFINITIONMAIN2015} that we adopt here.
We illustrate this Fig.~\ref{fig:mstar_fmol_altMS}, where we plot the same $\fmol$ vs. $M_*$ scaling relation for a sample selected using the offset from the main sequence as defined by \citet{saintongeColdInterstellarMedium2022} and recover the same drop-off beyond $M_*\sim10^{10.5}\,\msun$ they report, {\bx while the scaling relations at the low-mass end do not change significantly}.
We discuss the impact of the choice of main sequence definition in more detail in Appendix \ref{app:sf_criteria}, where we also show that the inclusion of starburst galaxies in our sample does not significantly change the molecular gas scaling relations we obtain.
\citet{caletteHIH2TOSTELLARMASS2018} adopt a morphology-based criterion to divide their sample into late and early types, so that neither of their subsamples is directly comparable to our star-forming sample.
They obtain a double power-law scaling relation from fitting to the means of binned data calculated using Kaplan-Meier estimators.
At low masses ($M_* \ll 1.74\times 10^{9}\,\msun$) their best fit $\fmol$ vs. $M_*$ slope for both late and early types agrees with what we find here, but at higher masses it is significantly steeper (see Fig.~\ref{fig:mstar_fmol}).
This is not surprising since at low masses, star-forming galaxies dominate the population, while at higher masses, the morphology based selection is likely to include a significant fraction of passive quenched galaxies.
When we adopt a main-sequence definition with a high mass turnover, we find a similar steepening trend as reported by \citet{caletteHIH2TOSTELLARMASS2018} as can be seen in Fig.~\ref{fig:mstar_fmol_altMS}.

In the low-mass regime ($M_*<10^{9}\,\msun$), observations of molecular gas are much more challenging, and scaling relations are less well constrained as a consequence.
One recent study \citep{huntScalingRelationsBaryonic2020} reports a single power-law scaling relation with a negative slope for $\fmol$ across a wide mass range of stellar masses ($10^{7}\,\msun<M_*<10^{11}\,\msun$).
This stands in contrast to our results, which indicate that the flat trend observed in the range $10^{9}\,\msun<M_*<10^{10}\,\msun$ extends down to $M_*=10^{8.5}\,\msun$.
However, while our sample includes non-detections as upper limits, the data analyzed in \citet{huntScalingRelationsBaryonic2020} is limited to CO detections, which is likely to introduce a bias towards extraordinarily gas-rich objects at the low-mass end ($10^{7}\,\msun<M_*<10^{9}\,\msun$).
This could explain the increasing molecular gas fraction they observe at these masses.
The color scale in Fig.~\ref{fig:mstar_fmol} additionally illustrates that the scatter in the molecular gas fraction is correlated with specific star formation rate (sSFR).
Galaxies found above the best-fit relation tend to have a higher sSFR, while those below tend to be less star forming.
A positive correlation between $\fmol$ and sSFR is also reported in the literature \citep{saintongeXCOLDGASSComplete2017, saintongeCOLDGASSIRAM2011} and expected, as this relation can be seen as a variation of the SK-law (see \citealt{saintongeColdInterstellarMedium2022} for a detailed review).
The physical origin of the scatter in $\fmol$ is not clear however.
The environment in which galaxies evolve could be a factor in this, as it is a parameter independent of galaxy mass.
We explore this possibility in Sect.~\ref{sec:env}.

\subsection{Molecular gas depletion time}
Next to the overall availability of molecular gas, the rate of star formation also depends on the efficiency with which clouds convert gas mass into stars.
While it is difficult to determine the true efficiency (i.e., how much of the available molecular gas will ultimately be converted into stars), we can measure the timescale of star formation from molecular gas, also called the molecular depletion time ($\tmol=\mmol/\mathrm{SFR}$), which is the reciprocal of the star formation efficiency ($\mathrm{SFE=SFR}/\mmol$).
\begin{figure}[htb]
\centering
    \includegraphics[clip=true, trim=0.25cm 0cm 0.1cm 0.1cm, width=\columnwidth]{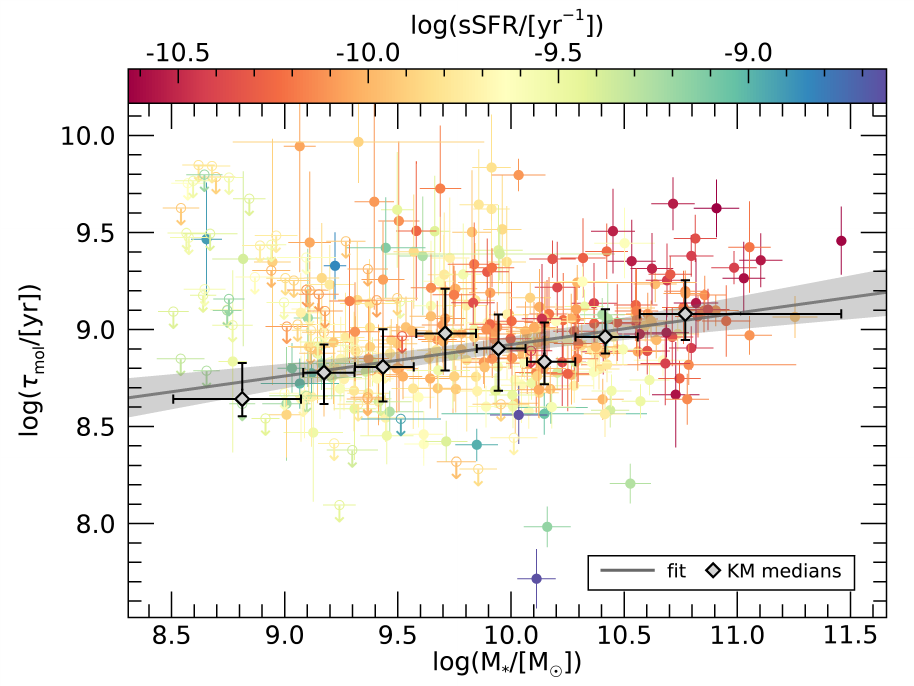}
     \caption{Molecular gas depletion time $\tmol = \mmol/\mathrm{SFR}$ vs. stellar mass. The thick dashed line shows the best fit from the Bayesian linear regression {\bx to the unbinned data} with the shaded area indicating the 95\% confidence band. The black diamonds correspond to the binned medians computed using the Kaplan-Meier estimator. {\bx The associated error bars show the extent of the bin (in $x$) and the 25-75 percentile interval (in $y$)}. Open symbols with downward arrows denote upper limits.}
   \label{fig:tdep}
\end{figure}

Figure~\ref{fig:tdep} shows $\tmol$ plotted against stellar mass for our star-forming sample.
Unlike $\fmol$, $\tmol$ shows some residual dependence on stellar mass despite being an intensive galaxy property.
While this correlation is much weaker than the $\mmol$ vs. $M_*$ and $\mmol$ vs. $\sfr$ relations (see Table~\ref{tab:fit_params}), {\bx its best-fit slope is non-zero at the 3$\sigma$ level.}
This is corroborated by the Kaplan-Meier estimated medians {\bx and 25-75 percentile intervals, which overlap with the fit to the unbinned data at all masses.}
Such a mass scaling indicates that higher mass galaxies consume their molecular gas reservoir through star formation less rapidly than their low-mass counterparts, despite their similar $\fmol$ values.
There seems to be no clear consensus in the literature concerning a possible correlation between molecular gas depletion time and stellar mass.
Some authors report a mild dependence of $\tmol$ on stellar mass in the mass range we investigate here \citep{boselliColdGasProperties2014, saintongeXCOLDGASSComplete2017, huntScalingRelationsBaryonic2020}, while others see no statistically significant correlation \citep{accursoDerivingMultivariateACO2017}.
A possible explanation for the dependence of $\tmol$ on stellar mass are properties of molecular clouds being affected by global galaxy properties \citep{sunMolecularGasProperties2020,rosolowskyGiantMolecularCloud2021} resulting in galaxies following different Kennicutt-Schmidt relations \citep{ellisonALMaQUESTSurveyNonuniversality2021}.
In particular the increased presence of stellar bulges in high mass galaxies has been linked to higher depletion times \citep{saintongeCOLDGASSIRAM2011, davisATLAS3DProjectXXVIII2014} and it has been suggested that their gravitational potential could stabilize the gas disk against fragmentation and collapse into star-forming clumps \citep{martigMORPHOLOGICALQUENCHINGSTAR2009}.
This could explain how specific star formation rate decreases with increasing stellar mass, despite the flat molecular gas fraction we find here.
\subsection{Neutral atomic gas mass}
\begin{figure}[htb]
\centering
	\includegraphics[clip=true, trim=0.25cm 0cm 0.1cm 0.1cm, width=\columnwidth]{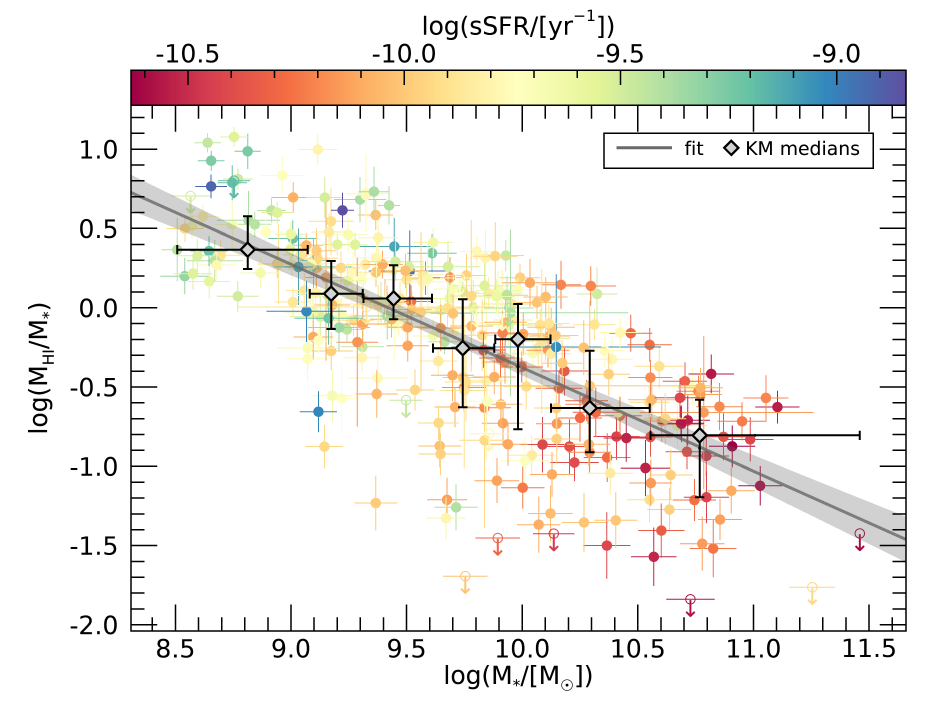}
     \caption{Neutral atomic gas mass fraction vs. stellar mass for the star-forming sample. The thick dashed line shows the best fit from the Bayesian linear regression {\bx to the unbinned data} with the shaded area indicating the 95\% confidence band. The black diamonds correspond to the binned medians computed using the Kaplan-Meier estimator. {\bx The associated error bars show the extent of the bin (in $x$) and the 25-75 percentile interval (in $y$)}. Open symbols with downward arrows denote upper limits.}
   \label{fig:mstar_fmhi}
\end{figure}

Since almost all galaxies in our sample have constraints on their neutral atomic gas mass from archival HI~21\,cm data, we can compare the mass scaling of the neutral atomic gas fraction $\fmhi=M_\mathrm{HI}/M_*$ to that of the molecular gas fraction $\fmol$.
Figure~\ref{fig:mstar_fmhi} shows $\fmhi$ plotted against stellar mass with a power-law fit from Bayesian linear regression and binned medians from survival analysis analogous to the $\fmol$ plots above.
We find that, contrary to $\fmol$, $\fmhi$ is anticorrelated with stellar mass, and that a single power-law is a good fit to our sample (see Table~\ref{tab:fit_params}).

The anticorrelation of $\fmhi$ and $M_*$ we find generally aligns with results from previous studies of similar samples \citep{saintongeXCOLDGASSComplete2017, caletteHIH2TOSTELLARMASS2018, huntScalingRelationsBaryonic2020}.
Some authors report a ``break'' or transition in the relation at a certain mass with a steeper relation above that characteristic mass \citep[e.g.,][although the latter note that both a single and a double power law fit the data well]{huntScalingRelationsBaryonic2020, caletteHIH2TOSTELLARMASS2018}.
We find no evidence for such a transition mass in our data, the binned medians agree well with a single-slope power law.
\section{Role of the (group) environment}
\label{sec:env}
In order to investigate environmental effects on the cold gas content of galaxies, we use information from the SDSS galaxy group catalog by \citet{tinkerSelfCalibratingHaloBasedGroup2021}.
This allows us to split our sample into galaxies that reside in groups or clusters and those that are isolated.
The non-isolated galaxies are further split into centrals and satellites as described in Sect.~\ref{sec:env_methods}.

\subsection{Molecular gas}
\begin{figure}[htb]
\centering
	\includegraphics[clip=true, trim=0.2cm 0.cm 0.1cm 1.2cm, width=\columnwidth]{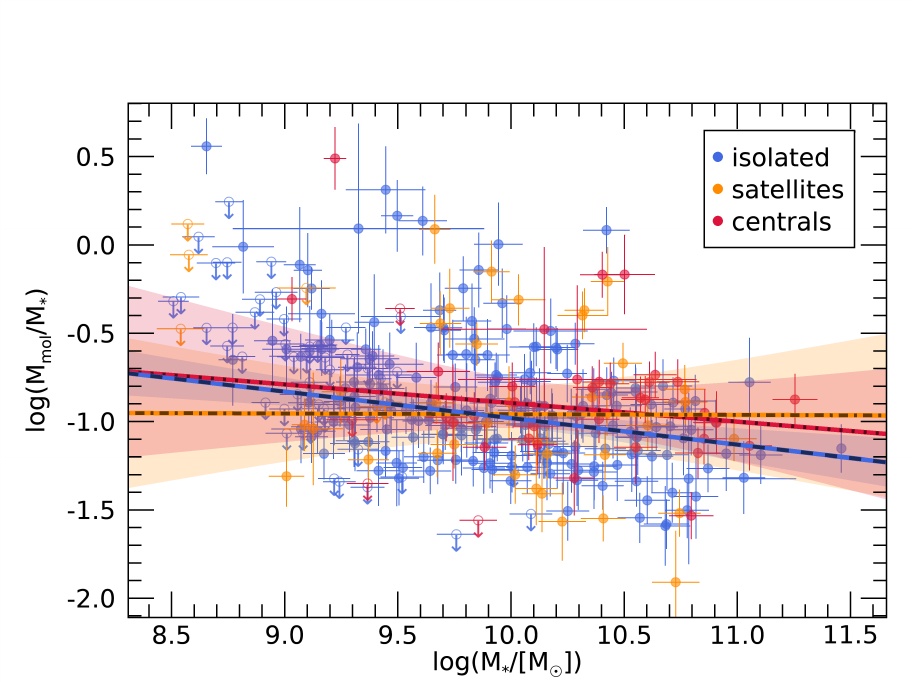}
    \includegraphics[clip=true, trim=0.2cm 0.cm 0.1cm 1.2cm, width=\columnwidth]{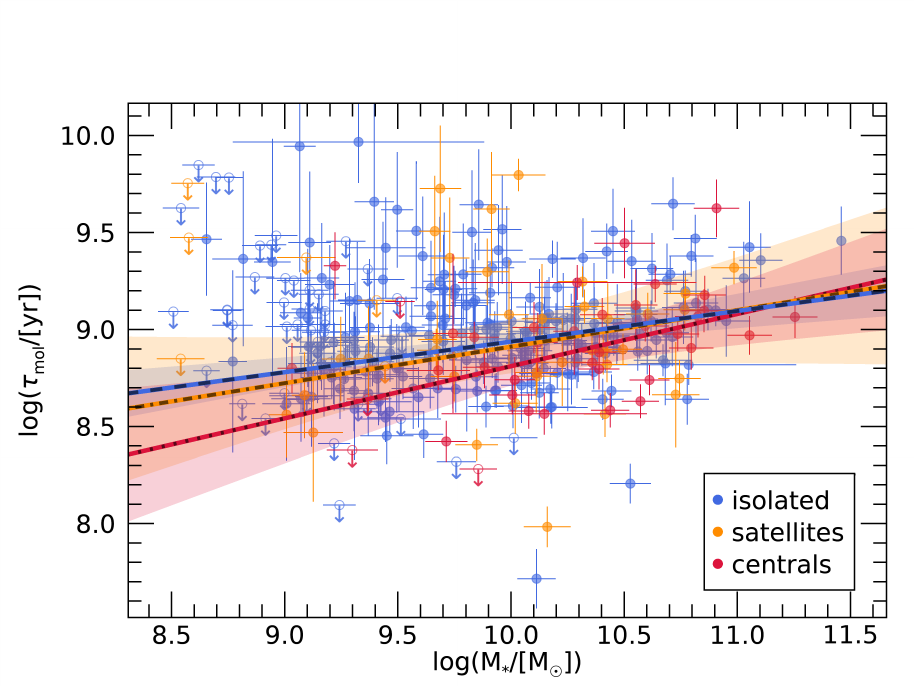}
     \caption{Molecular gas scaling relations by environment. \textit{Top:} Molecular gas mass fraction vs. stellar mass by environment. \textit{Bottom:} Molecular gas depletion time vs. stellar mass by environment. The environment is classified as central (red), satellite (orange), and isolated (blue). The best fits from the Bayesian linear regression {\bx to the unbinned data} are indicated by the blue-black-dashed (isolated), red-black-dotted (centrals), and orange-black-dash-dotted (satellites) lines respectively. The shaded areas indicate the 95\% confidence bands. Open symbols with downward arrows denote upper limits. We note that for a few galaxies there was no environment information available from the group catalog.}
   \label{fig:fmol_tdep_env}
\end{figure}

The top panel in Fig.~\ref{fig:fmol_tdep_env} shows the molecular gas fraction plotted against the stellar mass but with the sample split into the three environment classes described above.
The Bayesian linear regression performed on each subsample yield best-fit slopes that are consistent with each other and with the complete star forming sample within their respective $1\sigma$ errors.
Details on the best-fit parameters are listed in Table~\ref{tab:fit_env}.
There is no significant correlation between $M_*$ and $\fmol$ in any of the subsamples, nor is there a significant offset between the samples.
\begin{table}[htb]
    \caption{Fit results for environment subsamples.}
    \centering
    \begin{tabular}{lcccc}
    \hline
    \hline
    $f_\mathrm{mol}$ vs. $M_*$ & & & & \\
    \hline
    sample & intercept & slope & $\sigma$ & $\rho_{corr}$\\
    \hline
       all & $\hphantom{-}0.13\pm  0.38$ & $-0.11\pm  0.04$ & $ 0.36$ & $-0.19$\\
  isolated & $\hphantom{-}0.52\pm  0.42$ & $-0.15\pm  0.04$ & $ 0.33$ & $-0.26$\\
 satellite & $-0.92\pm  1.45$ & $-0.00\pm  0.14$ & $ 0.45$ & $-0.01$\\
   central & $\hphantom{-}0.16\pm  1.43$ & $-0.11\pm  0.14$ & $ 0.40$ & $-0.16$\\
   \hline
    $\tau_\mathrm{mol}$ vs. $M_*$ & & & & \\
    \hline
    sample & intercept & slope & $\sigma$ & $\rho_{corr}$\\
    \hline
       all & $\hphantom{-}7.29\pm  0.32$ & $\hphantom{-}0.16\pm  0.03$ & $ 0.28$ & $\hphantom{-}0.33$\\
  isolated & $\hphantom{-}7.36\pm  0.37$ & $\hphantom{-}0.16\pm  0.04$ & $ 0.27$ & $\hphantom{-}0.33$\\
 satellite & $\hphantom{-}7.03\pm  1.22$ & $\hphantom{-}0.19\pm  0.12$ & $ 0.37$ & $\hphantom{-}0.32$\\
   central & $\hphantom{-}6.12\pm  0.96$ & $\hphantom{-}0.27\pm  0.09$ & $ 0.26$ & $\hphantom{-}0.53$\\
   \hline
   $f_\mathrm{HI}$ vs. $M_*$ & & & & \\
    \hline
    sample & intercept & slope & $\sigma$ & $\rho_{corr}$\\
    \hline
       all & $\hphantom{-}6.16\pm  0.36$ & $-0.65\pm  0.04$ & $ 0.38$ & $-0.73$\\
  isolated & $\hphantom{-}6.67\pm  0.41$ & $-0.71\pm  0.04$ & $ 0.36$ & $-0.76$\\
 satellite & $\hphantom{-}4.38\pm  1.26$ & $-0.48\pm  0.13$ & $ 0.45$ & $-0.61$\\
   central & $\hphantom{-}5.81\pm  1.91$ & $-0.61\pm  0.19$ & $ 0.54$ & $-0.62$\\
   \hline
    \end{tabular}
    \caption*{Notes: Values quoted are best-fit parameters and $1\sigma$ uncertainties from the Bayesian linear fit to the $f_\mathrm{mol}$ vs. $M_*$, $\tau_\mathrm{mol}$ vs. $M_*$, and $f_\mathrm{HI}$ vs. $M_*$ relations in the isolated, satellite, and central subsamples. Also listed are the intrinsic scatter $\sigma$ and the linear correlation coefficient $\rho_\mathrm{corr}$.}
    \label{tab:fit_env}
\end{table}

{\bx The largest discrepancy is between isolated and satellite galaxies.
For the isolated galaxies, the best-fit slope is steeper than for the full sample, and this is the only subsample for which it is not within $3\sigma$ of zero.
The satellite subsample on the other hand has the shallowest best-fit slope.
However, taking into account the uncertainties on the best fit parameters, both subsamples are consistent with the best fit-relation obtained for the full star-forming sample.}
The bottom panel in Fig.~\ref{fig:fmol_tdep_env} shows the molecular gas depletion time versus stellar mass plot with the sample divided into the same sub-categories of isolated, satellite, and central galaxies.
Once again the best-fit parameters for the isolated, satellite, and central subsamples remain consistent with those obtained for the entire star-forming sample to within the estimated uncertainties, and no significant variations are observed as a function of environment.
\subsection{Neutral atomic gas}
\begin{figure}[htb]
\centering
	\includegraphics[clip=true, trim=0.2cm 0.cm 0.1cm 1.2cm, width=\columnwidth]{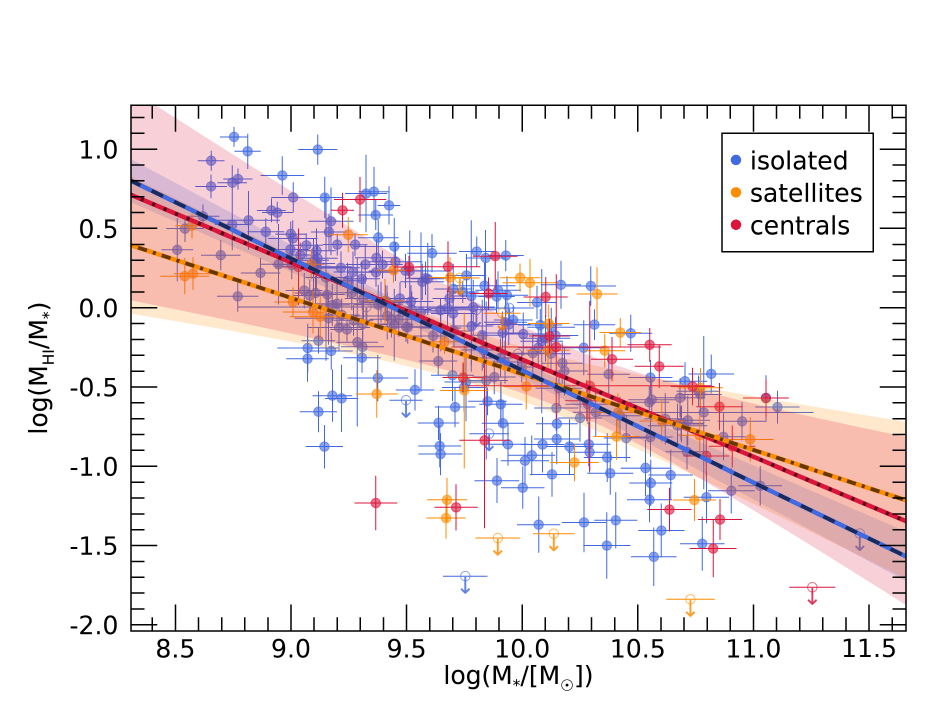}
     \caption{Neutral atomic gas fraction vs. stellar mass for central galaxies (red), satellites (orange), and isolated galaxies (blue).}
   \label{fig:fmhi_env}
\end{figure}

Neutral atomic hydrogen resides at much lower densities and larger radii in galaxies than its molecular counterpart, and should thus be more readily influenced by the environment.
The environmental impact on HI are well documented in the case of cluster galaxies, which have been found to be HI deficient \citep[e.g.,][]{HaynesInfluenceOfEnvironment1984}, but the picture for group galaxies is less clear \citep{appletonSHOCKENHANCEDEMISSIONDETECTION2013a, appletonMultiphaseGasInteractions2023a, deblokHighresolutionMosaicNeutral2018}.
In Fig.~\ref{fig:fmhi_env} we show the neutral atomic gas fraction $\fmhi$ plotted against stellar mass for our three environment based subsamples.
Much like in the case of molecular gas, the three separate Bayesian linear regression fits remain consistent with the results for the entire star-forming sample.

\section{Discussion and conclusions}
\label{sec:conclusion}
In our reanalysis of the combined ALLSMOG and xCOLDGASS star-forming galaxies, we include CO non-detections as 3$\sigma$ upper limits. 
We apply a metallicity and radiation field dependent $\alpha_\mathrm{CO}$ conversion factor that accounts for the dominant effect of turbulence in starburst galaxies \citep{accursoDerivingMultivariateACO2017}.
Our results indicate that the molecular gas fraction does not correlate with stellar mass for star-forming galaxies across the mass range $10^{8.5}\,\msun < M_* < 10^{10.5}\,\msun$.
For $M_*>10^{10.5}\,M_{\odot}$, the $\fmol$ vs. $M_{*}$ relation becomes highly sensitive to sample selection criteria, as evidenced by the comparison of our results with previous studies \citep{saintongeXCOLDGASSComplete2017, caletteHIH2TOSTELLARMASS2018} and our own tests (Appendix \ref{app:sf_criteria}).
We adopt a criterion based on the offset from the star-forming main sequence (following the definition of \citealt{renziniOBJECTIVEDEFINITIONMAIN2015}) in this work, and find the molecular gas fractions of the resulting sample are consistent with aforementioned flat trend for $M_* > 10^{10.5}\,\msun$.
Using a main sequence definition with a high-mass turnover instead \citep{saintongeColdInterstellarMedium2022} results in a sample with a sharp drop-off in the median molecular gas fraction above $M_*\sim10^{10.5}\,\msun$.
From this evidence we conclude that, in the local star-forming population in the mass range of $10^{8.5}\,\msun < M_* < 10^{10.5}\,\msun$, the abundance of cold molecular gas does not depend on stellar mass.
The molecular gas depletion time $\tmol$ on the other hand shows a mild positive dependence on stellar mass, indicating that more massive galaxies are less efficient at forming stars than their low-mass counterparts.
This is consistent with previous reports on gas consumption timescales \citep[e.g.,][]{boselliColdGasProperties2014, saintongeXCOLDGASSComplete2017, huntScalingRelationsBaryonic2020} and has been linked to the presence of stellar bulges \citep{saintongeCOLDGASSIRAM2011, davisATLAS3DProjectXXVIII2014} which could stabilize the gas against collapse \citep{martigMORPHOLOGICALQUENCHINGSTAR2009}.
In contrast to our results, an anticorrelation of $M_*$ and $\fmol$ has been reported in previous work on low-mass galaxies \citep[e.g.,][]{huntScalingRelationsBaryonic2020}.
This discrepancy between scaling relations in the literature is likely influenced by differences in methodology, particularly sample selection and assumptions on the $\alpha_\mathrm{CO}$.
Results differ in particular between studies based only on CO detections \citep[e.g.,][]{huntScalingRelationsBaryonic2020}, and those taking non-detections into account as upper limits \citep[e.g.,][]{saintongeXCOLDGASSComplete2017, caletteHIH2TOSTELLARMASS2018} as is the case in this work.
Examples where authors report scaling relations derived for the same sample with both approaches show that this can lead to significant differences.
\citet{bakerMolecularGasMain2022} report a $\sim$20\% steeper best-fit slope for the $\mmol$ vs. $M_*$ relation in their sample when fitting only detections compared to when taking non-detections into account.
The inclusion of upper limits requires statistical methods suited to extract information from censored data.
We employ three such techniques here, namely Bayesian linear regression \citep{kellyAspectsMeasurementError2007}, survival analysis in the form of the Kaplan-Meier estimator \citep{KaplanMeier1958}, and an estimator for Kendall's tau statistic \citep{oakesConsistencyKendallTau2008}. 
For our sample, all three methods yield results that are consistent with one another.
{\bx We additionally test spectral stacking for galaxies with $M_*<10^{9}\,\msun$ where our data are dominated by non-detections (see Appendix \ref{app:apex_stacks}), and find the results to be consistent with the other methods.}
A detailed comparison between commonly employed approaches can be found in \citet{saintongeColdInterstellarMedium2022}.
Finally, we note that while the inclusion of non-detections is important to avoid biasing the sample and recovering the flat trend we observe, we need to increase the number of CO detections at $M_*<10^{9}\,\msun$ significantly to place stronger constraints on the $\fmol$ vs. $M_*$ scaling relations in the low-mass regime.
{\bxx This would also allow to distinguish between a flat $\fmol$ vs. $M_*$ relation and a downturn in the low-mass regime, which cannot be done with upper limits from non-detections.}
Such constraints would require deeper observations of their molecular gas content in the same unbiased manner that the original ALLSMOG and xCOLDGASS surveys employed but with a much more sensitive telescope, such as the Atacama Large Aperture Submillimeter Telescope (AtLAST, see e.g. Mroczkowski et al. in prep; \citealt{Mroczkowski2023AtLAST, Ramasawmy2022AtLAST, Klaassen2020AtLAST}).
Another important difference between studies is the choice of $\alpha_\mathrm{CO}$.
The prescription we adopt for this work from \cite{accursoDerivingMultivariateACO2017} attempts to take into account variations of the $\alpha_{CO}$ on gas-phase metallicity, local radiation field and dynamical state of the molecular gas.
Other studies account only for the dependence on metallicity but vary in functional form and effective slope of the dependence, especially at low metallicities \citep{huntScalingRelationsBaryonic2020, gloverRelationshipMolecularHydrogen2011, wolfireDARKMOLECULARGAS2010}.
\citet{bakerMolecularGasMain2022} forego the metallicity dependence altogether and apply a constant $\alpha_\mathrm{CO}$ value based on data from the Milky Way.
To assess the effect of the adopted conversion factor, we repeated our analysis using the prescription recommended by \citet{huntScalingRelationsBaryonic2020}, which has a shallow slope compared to most other works, and a constant value corresponding to what is observed in the Milky Way \citep{bolattoCOtoHConversionFactor2013}.
We find that a metallicity-dependent conversion factor is necessary to recover the flat $\fmol$ scaling we report, but that the exact functional form can vary without fundamentally changing our results.
In low-mass, low-metallicity galaxies, where much of the molecular gas may be ``CO-dark'' \citep[e.g.,][]{maddenTracingTotalMolecular2020}, the CO(1-0) line may not be a good tracer of the overall molecular gas mass at all.
Some models suggest that the connection between CO(1-0) emission and molecular gas mass becomes increasingly complicated in the low-metallicity regime, depending strongly on the ``clumpiness'' of the ISM in addition to metallicity and radiation field \citep{ramambasonModelingMolecularGas2023}.
Other tracers, such as [CII]158$\mu$m and rotational lines of [CI], have been suggested as more effective in recovering molecular gas masses in low-metallicity galaxies \citep{ramambasonModelingMolecularGas2023, papadopoulosSubthermalExcitationLines2021, bisbasPhotodissociationRegionDiagnostics2021}, and could be used instead of relying on increasingly complex prescriptions for $\alpha_\mathrm{CO}$.
However, these tracers are relatively more challenging to observe, in particular [CI], which can be observed from ground in local galaxies in both its transitions (at rest-frame frequencies of 492 GHz and 809 GHz), but these observations require a sensitive telescope placed on a high dry site with  optimal submillimeter weather conditions, such as AtLAST.

While we find the molecular gas fraction to be independent of stellar mass in local star-forming galaxies, there is significant scatter of 0.31\,dex around this relation.
This scatter correlates with sSFR and is consistent with the scatter in SFR observed around the main sequence.
We investigate the galaxy (group) environment as a possible factor in the origin of this scatter, by analyzing the differences between central, satellite, and isolated galaxies in our sample, classified using the SDSS group catalog by \citet{tinkerSelfCalibratingHaloBasedGroup2021}.
Our results indicate that the $\fmol$ vs. $M_*$, $\tau_\mathrm{mol}$ vs. $M_*$, and $f_\mathrm{HI}$ vs. $M_*$ scaling relations do not differ significantly between the three subsamples.
From this we conclude that the group environment is not likely to be the primary driver of the scatter around the constant molecular gas fraction we find in our sample.
It is possible that a more nuanced separation of galaxies by environment, using additional parameters such as group size or separation between members, could reveal environmental effects that our classification is not sensitive to, but we do not have the sample size to perform such an analysis here.

\begin{acknowledgements}
      {\bx We would like to thank the anonymous referee for their insightful comments.}
      This project has received funding from the European Union’s Horizon 2020 research and innovation programme under grant agreement No 951815 (AtLAST). 
      S. Shen acknowledges support from the European High Performance Computing Joint Undertaking (EuroHPC JU) and the Research Council of Norway through the funding of the SPACE Centre of Excellence (grant agreement No 101093441).
      C. Cicone, P. Severgnini, and C. Vignali acknowledge a financial contribution from the Bando Ricerca Fondamentale INAF 2022 Large Grant, ‘Dual and binary supermassive black holes in the multi-messenger era: from galaxy mergers to gravitational waves’.
      This publication is based on data acquired with the Atacama Pathfinder Experiment (APEX) under programme ID 192.A-0359. APEX is a collaboration between the Max-Planck-Institut f\"ur Radioastronomie, the European Southern Observatory, and the Onsala Space Observatory. The calibrated and reduced APEX spectra of the ALLSMOG survey are publicly available through the ESO Phase 3 archive (\href{http://archive.eso.org/wdb/wdb/adp/phase3_spectral/form?collection_name=ALLSMOG}{ALLSMOG Phase 3 data release}).
\end{acknowledgements}
\newpage
\clearpage

\bibliography{Project1_sources}

\appendix
\section{Impact of the star-forming sample selection}
\label{app:sf_criteria}
\begin{figure}[htb]
\centering
    \includegraphics[clip=true, trim=0.25cm 0cm 0.1cm 0.1cm, width=\columnwidth]{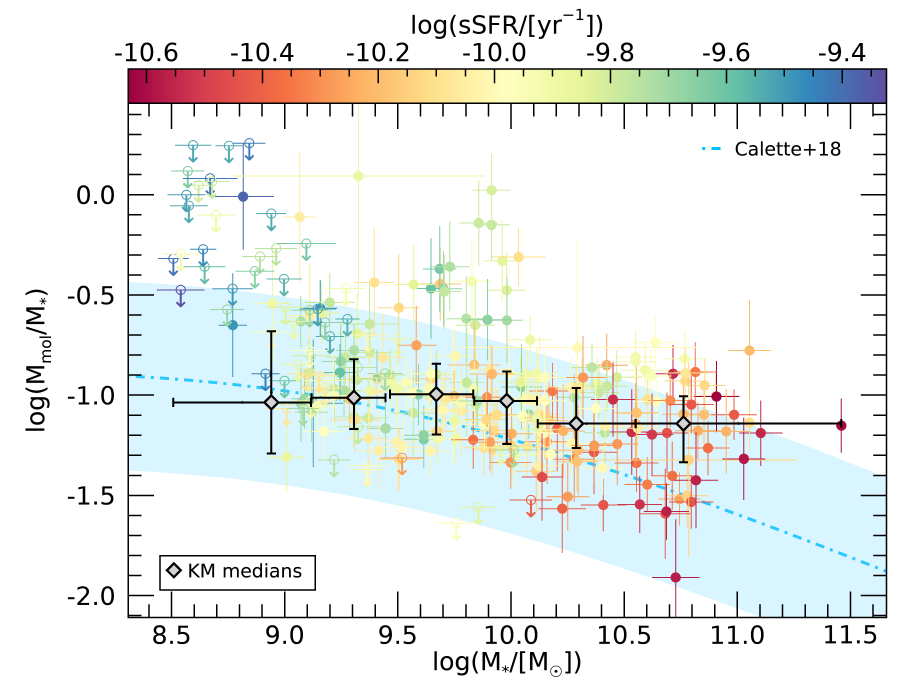}
    \includegraphics[clip=true, trim=0.25cm 0cm 0.1cm 0.1cm, width=\columnwidth]{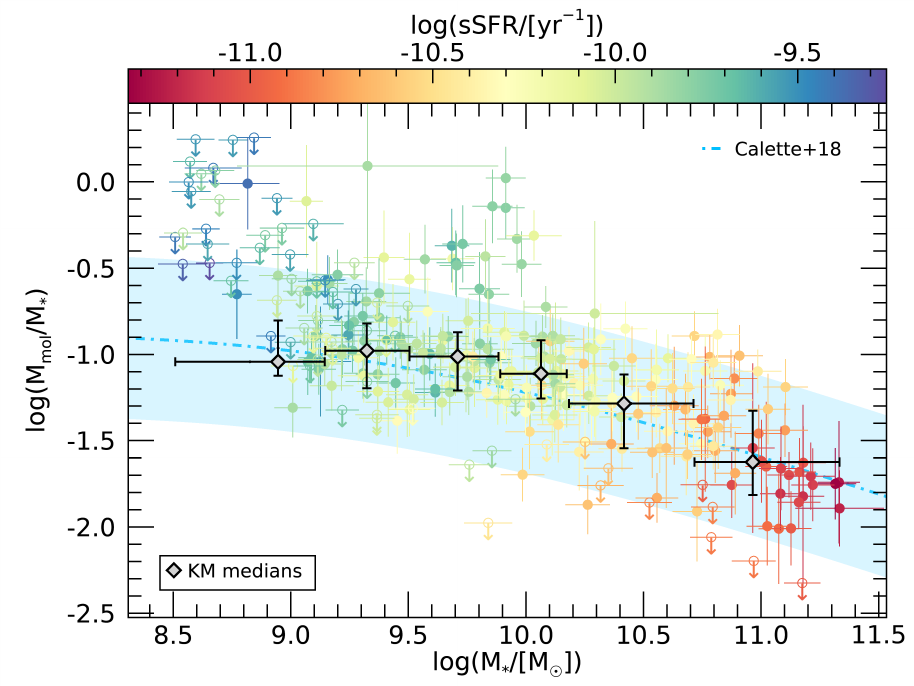}
     \caption{Molecular gas fraction $\fmol = \mmol/\mathrm{SFR}$ vs. stellar mass for main sequence galaxies according to the definitions of \citet{renziniOBJECTIVEDEFINITIONMAIN2015} ({\it top}) and \citet{saintongeColdInterstellarMedium2022} ({\it bottom}), respectively. The black diamonds correspond to the binned medians computed using the Kaplan-Meier estimator. Open symbols with downward arrows denote upper limits.}
   \label{fig:fmol_deltamscut}
\end{figure}
We explore the effect of the chosen criterion for classifying galaxies as star-forming on the resulting scaling relations by repeating the analysis with different criteria.
First, we test how the inclusion of starbursts in our sample impacts the $\fmol$ vs. $M_*$ scaling relation.
To this end, we change the criterion of $\log\Delta (\mathrm{MS}) > -0.4$\,dex to $|\log\Delta (\mathrm{MS})| < 0.4$\,dex while maintaining the definition of the main sequence following \citet{renziniOBJECTIVEDEFINITIONMAIN2015}.
This ensures that galaxies with specific star formation rates high above the main sequence are excluded from the sample, in addition to the passive quenched galaxies.
The $\fmol$ vs. $M_*$ relation for this altered sample is shown in the top panel of Fig.~\ref{fig:fmol_deltamscut}.
Comparing this to the main sequence and starburst sample from Fig.~\ref{fig:mstar_fmol} makes it clear that the inclusion of starbursts does not significantly change the overall scaling relation.
This is unsurprising insofar as the different physics of starbursts are, to some degree, taken into account in the $\alpha_\mathrm{CO}$ prescription we adopt in our analysis, as discussed in Sect.~\ref{sec:alpha_co_method}.
We perform the same test for the curved main sequence definition from \citet{saintongeColdInterstellarMedium2022} and find the same result (see bottom panel of Fig.~\ref{fig:fmol_deltamscut} compared to Fig.~\ref{fig:mstar_fmol_altMS}).
\begin{figure}[htb]
\centering
    \includegraphics[clip=true, trim=0.25cm 0cm 0.1cm 0.1cm, width=\columnwidth]{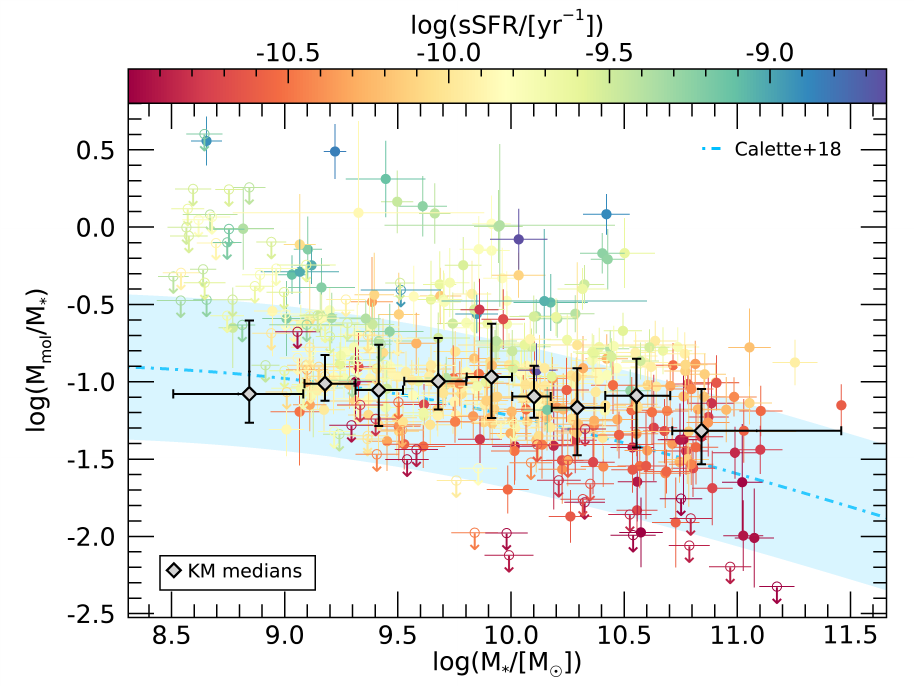}
     \caption{Molecular gas fraction $\fmol = \mmol/\mathrm{SFR}$ vs. stellar mass for the sSFR selected sample. The black diamonds correspond to the binned medians computed using the Kaplan-Meier estimator. {\bx The associated error bars show the extent of the bin (in $x$) and the 25-75 percentile interval (in $y$)}. Open symbols with downward arrows denote upper limits.}
   \label{fig:fmol_ssfrcut}
\end{figure}
\begin{figure}[htb]
\centering
    \includegraphics[clip=true, trim=0.25cm 0cm 0.1cm 0.1cm, width=\columnwidth]{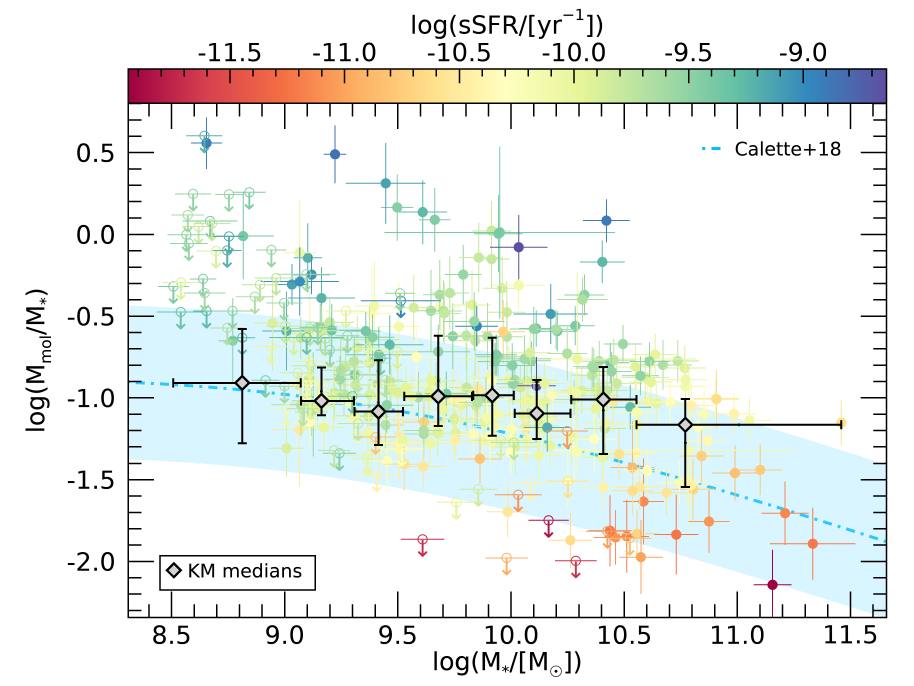}
     \caption{Molecular gas fraction $\fmol = \mmol/\mathrm{SFR}$ vs. stellar mass for the BPT-selected SF plus composite sample. The black diamonds correspond to the binned medians computed using the Kaplan-Meier estimator. {\bx The associated error bars show the extent of the bin (in $x$) and the 25-75 percentile interval (in $y$)}. {\bx The associated error bars show the extent of the bin (in $x$) and the 25-75 percentile interval (in $y$)}. Open symbols with downward arrows denote upper limits.}
   \label{fig:fmol_bptcut}
\end{figure}

Second, we test approaches to sample selection which do not rely on an analytic main sequence definition.
One such approach is to select galaxies that have $\log\mathrm{(sSFR/[yr^{-1}])} > -11$ regardless of stellar mass.
The results, shown in Fig.~\ref{fig:fmol_ssfrcut}, are similar to those obtained when assuming the \citet{renziniOBJECTIVEDEFINITIONMAIN2015} main sequence (Fig. \ref{fig:mstar_fmol}).
Another frequently used selection method (for example in \citealt{bakerMolecularGasMain2022}) relies on the BPT classification (see Sect. \ref{sec:bpt}).
Using the BPT diagram, we exclude AGN and use only galaxies classified as SF or composite as our sample.
The results, shown in Fig. \ref{fig:fmol_bptcut}, are again close to those of the linear main sequence and constant sSFR cutoff samples.
We note that, while the choice of how to demarcate the star-forming population clearly affects the scaling relation between $M_*$ and $\fmol$ at high masses (above $M_*\sim10^{10.5}\,\msun$), there is no significant effect on the low-mass regime.
Our predictions for the molecular gas scaling relations in the very low-mass regime ($M_* < 10^{9}\,\msun$ should therefore be robust against differences in sample selection.

\section{Choice of CO-to-H$_2$ conversion factor}
\label{app:alphaco}
To test the influence of the chosen prescription for the CO-to-H$_2$ conversion factor $\alpha_\mathrm{CO}$ on our results, we re-derive the $\fmol$ vs. $M_*$ scaling relation using two additional prescriptions.
The first is the simple assumption of a constant $\alpha_\mathrm{CO}$ with a Milky Way derived value of $4.35\,\acounits$ \citep{bolattoCOtoHConversionFactor2013}.
The second is a prescription proposed by \citet{huntScalingRelationsBaryonic2020} based on a sample of local low-mass ($M_*\sim 10^7-10^{11}\,\msun$) star-forming galaxies.
It follows a simple power law scaling with gas-phase metallicity for low-metallicity galaxies while keeping $\alpha_\mathrm{CO}$ constant in the high-metallicity regime:
\begin{equation}
\alpha_\mathrm{CO} = 
\left\{
	\begin{array}{ll}
		\alpha_\mathrm{CO,\odot}(Z/Z_\odot)^{-1.55} & \mbox{if } Z<Z_\odot \\
		\alpha_\mathrm{CO,\odot} & \mbox{if } Z\geq Z_\odot.
	\end{array}
\right.
\end{equation}
Here $\alpha_\mathrm{CO,\odot}=4.35\,\acounits$ is the same Milky Way derived conversion factor as above and $Z_\odot=8.69$ is the solar metallicity in units of $12+\log\mathrm{(O/H)}$.
This prescription is functionally similar to previous suggestions for a metallicity dependent $\alpha_\mathrm{CO}$ \citep{gloverRelationshipMolecularHydrogen2011,wolfireDARKMOLECULARGAS2010} but has a somewhat shallower slope than most (see also \citealt{bolattoCOtoHConversionFactor2013} for a comparison between prescriptions).
Adopting the metallicity dependent prescription from \citet{huntScalingRelationsBaryonic2020} decreases the correlation between $M_*$ and $\fmol$ and boosts the correlation found between $M_*$ and $\tmol$ somewhat, but the overall picture remains unchanged.
The best-fit slopes for both relations are consistent with those obtained using the \citet{accursoDerivingMultivariateACO2017} factor, to within $3\sigma$.
Adopting a constant conversion factor instead results in a positive slope in the $\fmol$ vs. $M_*$ relation, as it systematically underestimates the molecular gas masses for low-mass, low-metallicity galaxies.
For the same reason it results in a steeper slope for the $\tau_\mathrm{mol}$ vs. $M_*$ relation.

\begin{table}[htb]
    \caption{Comparison between $\alpha_\mathrm{CO}$ prescriptions.}
    \centering
    \begin{tabular}{lcccc}
    $f_\mathrm{mol}$ vs. $M_*$ & & & & \\
    \hline
    \hline
    $\alpha_\mathrm{CO}$ & intercept & slope & $\sigma$ & $\rho_{corr}$\\
    \hline
    A17 & $\hphantom{-}0.13\pm  0.38$ & $-0.11\pm  0.04$ & $ 0.36$ & $-0.19$\\
    H20 & $-1.09\pm  0.35$ & $\hphantom{-}0.02\pm  0.04$ & $ 0.33$ & $\hphantom{-}0.03$\\
    MW & $-1.99\pm  0.34$ & $\hphantom{-}0.10\pm  0.03$ & $ 0.32$ & $\hphantom{-}0.19$\\
   \hline
    $\tau_\mathrm{mol}$ vs. $M_*$ & & & & \\
    \hline
    \hline
    $\alpha_\mathrm{CO}$ & intercept & slope & $\sigma$ & $\rho_{corr}$\\
    \hline
    A17 & $\hphantom{-}7.29\pm  0.32$ & $\hphantom{-}0.16\pm  0.03$ & $ 0.28$ & $\hphantom{-}0.33$\\
    H20 & $\hphantom{-}5.90\pm  0.30$ & $\hphantom{-}0.30\pm  0.03$ & $ 0.26$ & $\hphantom{-}0.58$\\
    MW & $\hphantom{-}4.91\pm  0.31$ & $\hphantom{-}0.40\pm  0.03$ & $ 0.27$ & $\hphantom{-}0.67$\\
   \hline
    \end{tabular}
    \caption*{Notes: Values quoted are best-fit parameters and $1\sigma$ uncertainties from the Bayesian linear fit to the $f_\mathrm{mol}$ vs. $M_*$ and $\tau_\mathrm{mol}$ vs. $M_*$ relations assuming three different $\alpha_\mathrm{CO}$ prescriptions according to \citet{accursoDerivingMultivariateACO2017} (A17) and \citet{huntScalingRelationsBaryonic2020} (H20), and a constant value corresponding to the one measured in the Milky Way (MW). Also listed are the intrinsic scatter $\sigma$ and the linear correlation coefficient $\rho_\mathrm{corr}$ for the respective fits.}
    \label{tab:aco_trends}
\end{table}

\section{Metallicity weighted fit}
\label{app:met_weights}
\begin{figure}[htb]
    \centering
    \includegraphics[clip=true, trim=0.65cm 0.22cm 1.cm 0.16cm, width=0.4\textwidth]{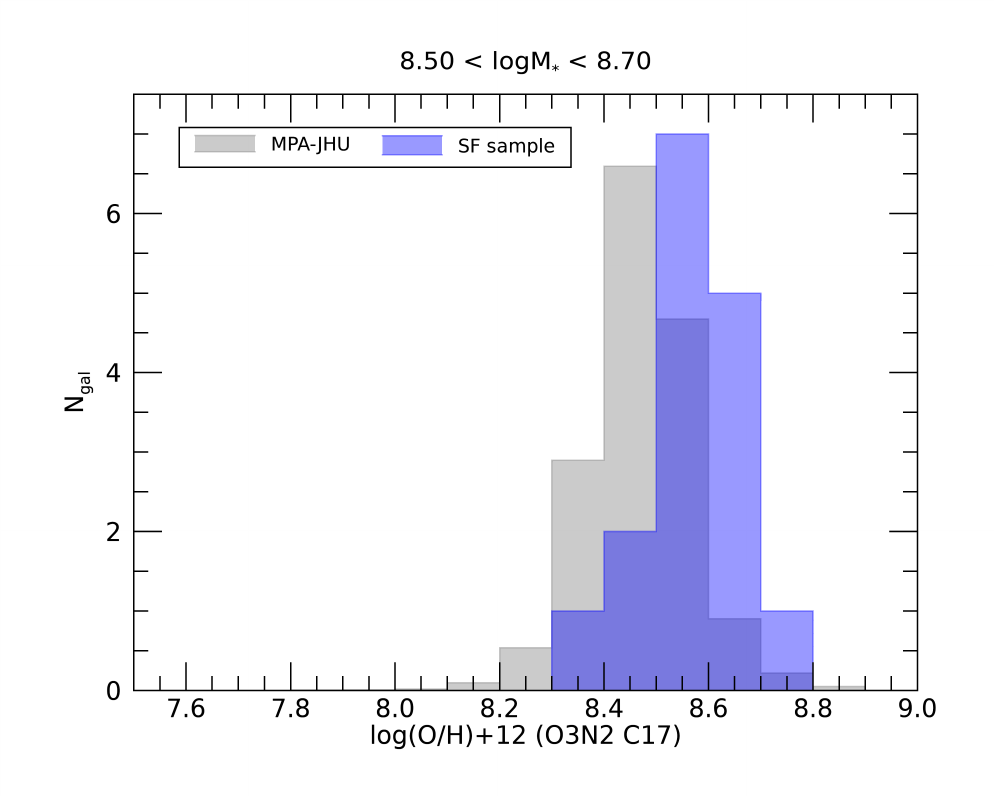}
    \includegraphics[clip=true, trim=0.65cm 0.22cm 1.cm 0.16cm, width=0.4\textwidth]{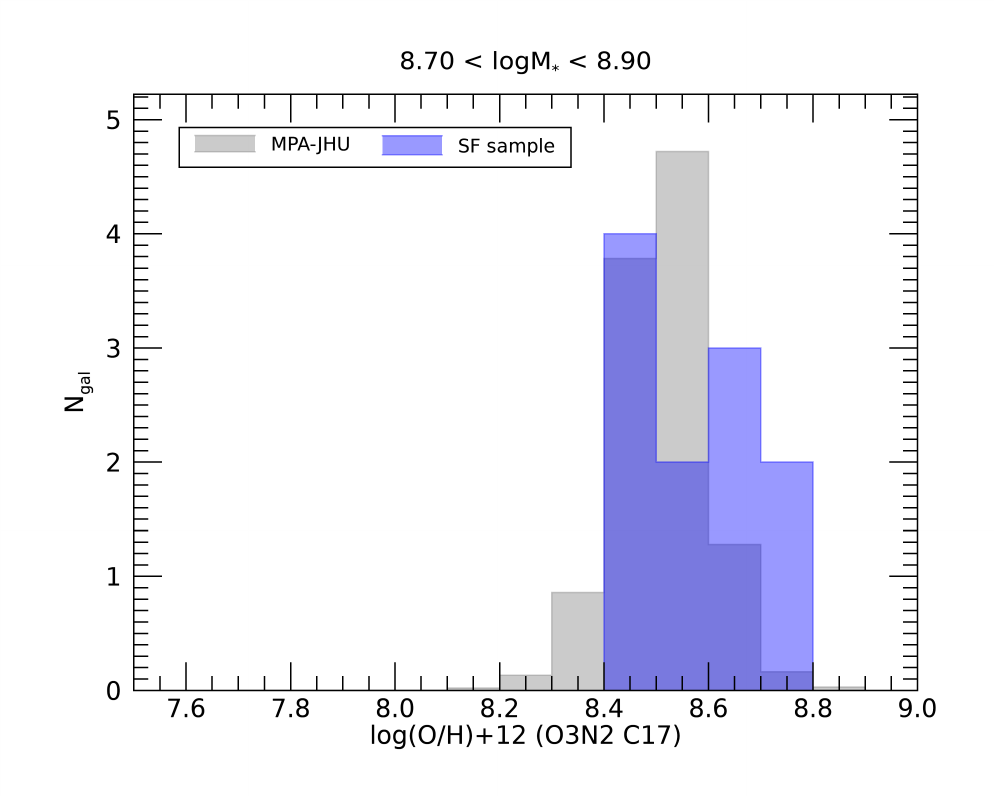}
    \includegraphics[clip=true, trim=0.65cm 0.22cm 1.cm 0.16cm, width=0.4\textwidth]{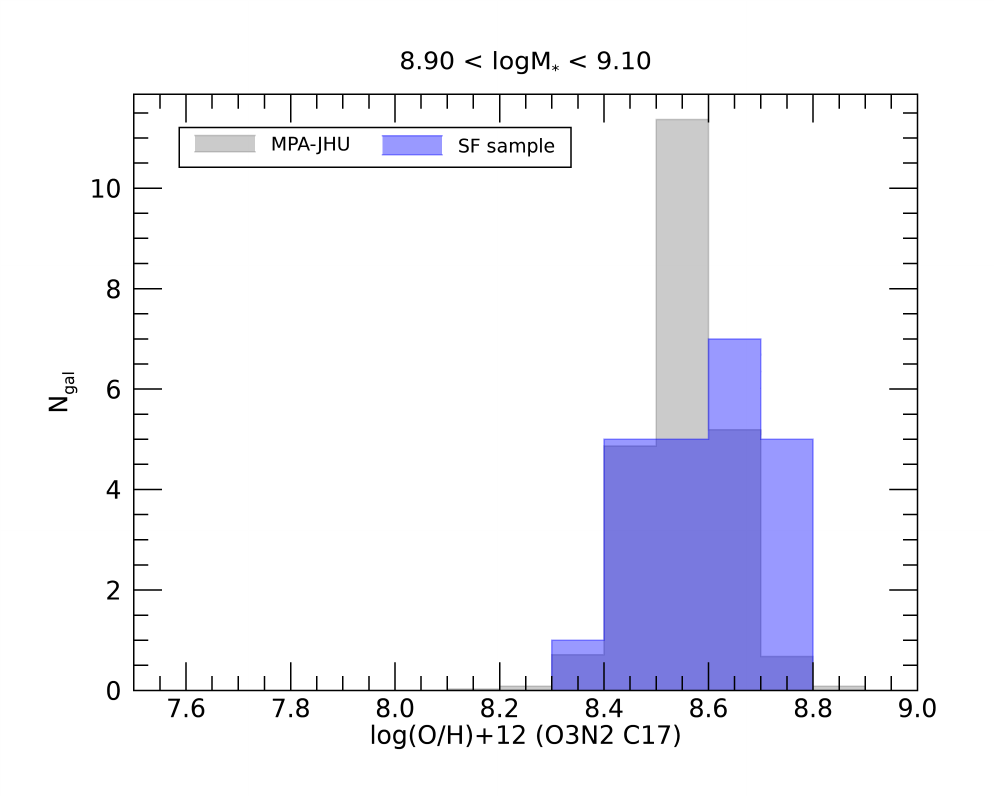}
    \caption{Gas-phase metallicity distribution of the full MPA-JHU star-forming sample vs. our sample in the three lowest stellar mass bins. The histograms are normalized to the real distribution in our sample for each bin.}
    \label{fig:met_hist_comp}
\end{figure}
At low masses ($M_* < 10^9\,\msun$), our sample is drawn entirely from the ALLSMOG survey.
Since the original survey selected galaxies above a certain gas phase metallicity (see Sect.~\ref{sec:sample} for details), we need to ensure our sample is not biased towards metal-rich galaxies in this mass range.
We do so by comparing the metallicity distribution of our sample against that of all star-forming galaxies from the full MPA-JHU catalog (selected using the same criterion of $\Delta\mathrm{(MS)}>-0.4$\,dex we outline in Sect.~\ref{sec:sample}).
For this comparison we bin both samples in stellar mass (with 0.2\,dex bin size) and metallicity (0.1\,dex bin size).
The results for the first three mass bins are shown in Fig.~\ref{fig:met_hist_comp}.
These examples illustrate that the distribution of gas-phase metallicity in our sample is indeed somewhat skewed towards high metallicities at low stellar masses.
In the $8.7<\log M_*/\msun < 8.9$ bin we achieve the worst coverage, but our sample range still covers just over 90\% of the MPA-JHU distribution.
To alleviate this bias, we calculate weights for each mass-metallicity bin by dividing the normalized histogram of the complete MPA-JHU sample by that of our sample.
During the fitting procedure, each galaxy in the mass range $10^{8.5}\,\msun < M_* < 10^{10.5}\,\msun$ where our sample shows signs of metallicity bias, is weighted by the appropriate factor for its bin.
\section{Stacking of low-mass non-detections}
\label{app:apex_stacks}
{\bx In the $8.5 < \log M_*/M_\odot < 9.0$ mass range, most of the data in our sample are upper limits.
While these upper limits are crucial in recovering unbiased scaling relations in the low-mass regime, they do not provide as strong a constraint as detections would.
Here, we test the possibility of improving the constraints derived from non-detections by spectral stacking.
Our sample contains 22 non-detections with stellar masses below $10^9\,M_\odot$ that were observed in the CO(2-1) emission line with APEX as part of the ALLSMOG survey.
The individual spectra of these non-detections are re-binned to a spectral resolution of 0.2\,km/s, because the native spectral resolution is not consistent between different observations.
Spectra are then co-added, weighted by their inverse variance (calculated over two spectral windows of [-900,-300]\,km/s and [300,900]\,km/s relative to the rest frequency of the CO(2-1) line inferred from the galaxies' redshifts), and re-binned again on a 10\,km/s grid.
The stacked spectrum shows no clear detection of the CO(2-1) line (see Fig.~\ref{fig:apex_stacks}).
Nonetheless, we can use the rms noise of the stacked spectrum to calculate an upper limit on the average CO(2-1) line flux of the galaxies in the bin following the procedure outlined in \cite{ciconeFinalDataRelease2017}.
{\bxx The resulting upper limit is about a factor of five lower than those calculated for the individual galaxies, since the rms noise scales with the inverse square root of the number of spectra in the stack when co-adding white noise.}
\begin{figure}[htb]
    \centering
    \includegraphics[clip=true, trim=0.65cm 0.22cm 1.cm 0.16cm, width=0.4\textwidth]{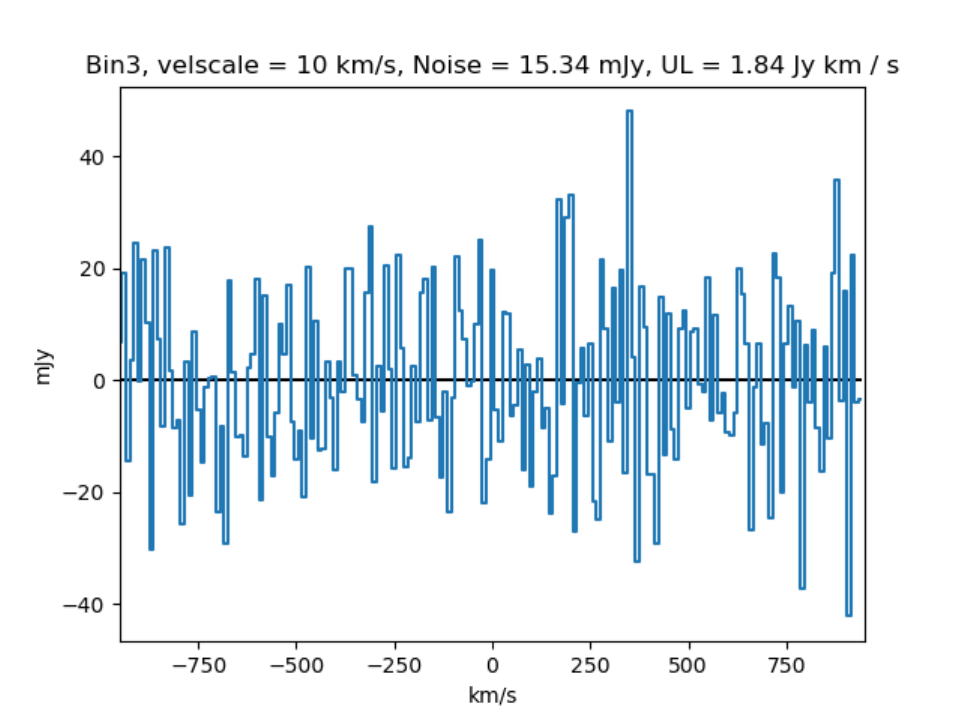}
    \caption{\bxx Stacked spectrum of 22 CO(2-1) non-detections with $M_*<10^9\,\msun$ from the ALLSMOG survey. The titles list the rms noise and the derived upper limits (UL) on the line flux.}
    \label{fig:apex_stacks}
\end{figure}

From the upper limit on the CO(2-1) flux, we calculate an upper limit on the mean molecular gas mass of the galaxies in the stack, following the same procedure as for the individual galaxies, as outlined in section \ref{sec:alpha_co_method}.
{\bxx For the stellar mass, luminosity distance, and redshift needed in the calculations, we use the mean of all galaxies in the stack.
The $\alpha_\mathrm{CO}$ conversion factor varies between $\sim3\,\acounits$ and $\sim20\,\acounits$ within the bin (with one notable outlier at $30\,\acounits$) and shows no trend with stellar mass.
We adopt a $\alpha_\mathrm{CO}=20\,\acounits$ in our computation, which corresponds to the higher end of the typical values in the bin, ensuring that the result is a meaningful upper limit.
The resulting upper limit is above, and therefore consistent with, the KM median of the lowest mass bin (see Fig.~\ref{fig:mstar_fmol}).}

\end{document}